\documentclass[11pt,american,twoside,a4paper]{article}
\usepackage[american,provide=*]{babel}
\usepackage[T1]{fontenc}
\usepackage[utf8]{inputenc}
\usepackage{a4wide}
\usepackage{fourier,bbm,bm}
\usepackage{amsmath,amsfonts,amsthm,amssymb}
\usepackage{latexsym,mathtools}
\usepackage{enumerate}
\usepackage{color,xcolor}
\usepackage{xargs,xspace,needspace,xparse,easyReview}
\usepackage[active]{srcltx}
\usepackage[colorlinks=true]{hyperref}
\usepackage{fancyhdr}
\colorlet{darkblue}{blue!50!black}
\hypersetup{
	colorlinks,%
	citecolor=darkblue,%
	filecolor=red,%
	linkcolor=darkblue,%
	urlcolor=darkblue,%
	pdfnewwindow=true,%
	pdfstartview={FitH}
}

\numberwithin{equation}{section}

\newcounter{smallarabics}

\newcounter{smallroman}

\newcommand{\ben}{\begin{enumerate}[{\rm (1)}]}
\newcommand{\een}{\end{enumerate}}

\newtheorem{theorem}{Theorem}[section]
\newtheorem{proposition}[theorem]{Proposition}
\newtheorem{lemma}[theorem]{Lemma}
\newtheorem{corollary}[theorem]{Corollary}
\newtheorem{definition}[theorem]{Definition}
\newtheorem{remark}[theorem]{Remark}

\newcommand{\beq}{\begin{equation}}
\newcommand{\eeq}{\end{equation}}
\newcommand{\bear}[1]{\begin{array}{#1}}
\newcommand{\ear}{\end{array}}
\newcommand{\bep}{\begin{proposition}}
\newcommand{\eep}{\end{proposition}}
\newcommand{\bet}{\begin{theoreme}}
\newcommand{\eet}{\end{theoreme}}
\newcommand{\bel}{\begin{lemma}}
\newcommand{\eel}{\end{lemma}}
\newcommand{\bed}{\begin{definition}}
\newcommand{\eed}{\end{definition}}
\newcommand{\ber}{\begin{remark}}
\newcommand{\eer}{\end{remark}}

\NewDocumentCommand{\ASS}{mm}{\expandafter\newcommand\csname #1\endcsname{{\hyperref[#1]{\bf (#2)}}}}
\NewDocumentCommand{\preASS}{mm}{\expandafter\newcommand\csname pre#1\endcsname{{\hyperref[#1]{\bf (#2)}}}}
\newcommand{\assuming}[3]{
\begin{quote}\label{#2}{\bf(#1) }%
#3%
\end{quote}%
\ASS{#2}{#1}}

\newcommand{\ds}{\displaystyle}

\newcommand{\ie}{\textsl{i.e.,\ }}

 \def\cB{\mathcal{B}} 
 \def\cE{\mathcal{E}} 
 \def\cH{\mathcal{H}} 
 \def\cK{\mathcal{K}} \def\cL{\mathcal{L}}
 \def\cN{\mathcal{N}} \def\cO{\mathcal{O}}

  \def\cX{\mathcal{X}}

\def\CC{\mathbb{C}}

\def\II{\mathbb{I}}

\def\RR{\mathbb{R}}

\def\one{\mathbb{1}}

\def\C{\mathbb{C}}

\def\R{\mathbb{R}}

\def\fA{\mathfrak{A}}

\def\fh{\mathfrak{h}}
\def\fH{\mathfrak{H}}

\def\fM{\mathfrak{M}}
\def\fF{\hbox{$\mathfrak{F}$}}

\def\fS{{\mathfrak S}}

\def\bF{\boldsymbol{F}}

\def\sS{\mathsf{S}}
\def\sR{\mathsf{R}}

\newcommand{\tr}{\mathop{\mathrm{tr}}\nolimits}
\newcommand{\bra}{\langle}
\newcommand{\ket}{\rangle}
\newcommand{\ran}{\mathop{\mathrm{Ran}}}
\renewcommand{\ker}{\mathop{\mathrm{Ker}}}
\newcommand{\Dom}{\mathop{\mathrm{Dom}}}
\renewcommand{\sp}{\mathop{\mathrm{sp}}}
\renewcommand{\Re}{\mathop{\mathrm{Re}}}
\renewcommand{\Im}{\mathop{\mathrm{Im}}}

\newcommand{\Tr}{\mathop{\mathrm{Tr}}}

\newcommand{\e}{\mathrm{e}}
\renewcommand{\d}{\mathrm{d}}
\renewcommand{\i}{\mathrm{i}}

\newcommand{\BAR}{\overline}

\renewcommand{\dim}{\mathop{\mathrm{dim}}}
\def\Ker{{\rm Ker}\,}
\def\interior{\mathop{\mathrm{int}}}
\def\closure{\mathop{\mathrm{cl}}}

\newcommand{\argdot}{\,{\bm\cdot}\,}

\newcommand{\CAR}{\mathop{\mathrm{CAR}}}

\newcommand{\ttm}{\mathrm{2tm}}
\newcommand{\anc}{\mathrm{east}}

\newcommand{\qpsc}{\mathrm{qpsc}}
\newcommand{\tm}{\mathrm{TM}}

\newcommand{\Ent}{\mathop{\mathrm{Ent}}}

\def\free{{\rm fr}}

\begin{document}
\def\today{}
\title{Entropic Fluctuations in Quantum Statistical Mechanics: A Review}
\author{T. Benoist$^{1}$, L. Bruneau$^{2}$, V. Jak\v{s}i\'c$^{3}$, A. Panati$^4$, C.-A. Pillet$^4$
\\ \\
$^1$ Universit\'e de Toulouse, INUC, UT2J, INSA Toulouse, TSE, CNRS, IMT, Toulouse, France\\
\texttt{tristan.benoist@math.univ-toulouse.fr}
\\ \\
$^2$ D\'epartement de Math\'ematiques, CNRS UMR 8088\\
Cergy Paris Universit\'e, 95000 Cergy-Pontoise, France\\
\texttt{laurent.bruneau@cyu.fr}
\\ \\
$^3$Dipartimento di Matematica, Politecnico di Milano\\
piazza Leonardo da Vinci 32, 20133 Milano,  Italy\\
\texttt{vojkan.jaksic@polimi.it}
\\ \\
$^4$Universit\'e de Toulon, CNRS, CPT, UMR 7332, 83957 La Garde, France\\
Aix-Marseille Univ, CNRS, CPT, UMR 7332, Case 907, 13288 Marseille, France\\
\texttt{panati@univ-tln.fr, pillet@univ-tln.fr}
}
\maketitle
\thispagestyle{empty}

\noindent{\small{\bf Abstract.} The celebrated Evans--Searles, respectively
Gallavotti--Cohen, fluctuation theorem concerns certain universal statistical
features of the entropy production rate of a classical system in a transient,
respectively steady, state. In this paper, we review and compare several
possible extensions of these fluctuation theorems to quantum systems. We present,
in an introductory way, the approach proposed in a recent series of works
\href{https://doi.org/10.1007/s11005-024-01777-0}{[Lett. Math. Phys. 114, 32 (2024)]}, \href{https://doi.org/10.1142/S0129055X24610063}{[Rev. Math. Phys 38, 2461006 (2026)]}, \href{https://doi.org/10.1007/s00220-025-05360-z}{[Commun. Math. Phys. 406, 201 (2025)]} 
and \href{https://doi.org/10.1063/5.0311182}{[J. Math. Phys. 67, 022101 (2026)]}, 
including a previously missing extension of the Gallavotti--Cohen (steady state)
version of the FT to two-time measurement entropy production in the algebraic
setting of quantum dynamical systems. We focus on the open quantum
systems, where a small system interacts with several (infinitely extended)
reservoirs, and discuss the following approaches: naive, two-time measurement,
phase space contraction and ancilla state tomography. We review the connection
with modular theory. The latter provides a way to extend the classical notions
to the quantum domain. We discuss the quantum version of the principle of
regular entropic fluctuations, introduced in the classical context in
\href{https://doi.org/10.1088/0951-7715/24/3/003}{[Nonlinearity 24, 699 (2011)]}. 
Finally, we relate the statistical properties of these various notions
of entropy production to spectral resonances of quantum transfer operators and
apply this approach to a paradigmatic model of non-equilibrium statistical
mechanics: the spin-fermion model. The results presented here shed a new light
on the nature of entropic fluctuations in quantum statistical mechanics.
}
\newpage
\tableofcontents

\section{Introduction}

Classical entropic fluctuation theorems (FT) have marked a turning point
in non-equilibrium statistical mechanics. Since their discovery in the
early 1990's~\cite{Evans1993,Evans1994,Gallavotti1995a}, a new area of
research has developed. The first numerical experiment~\cite{Evans1993}
were immediately followed by two separate sets of theoretical explanations
corresponding to different conceptual frameworks, by Evans and 
Searles~\cite{Evans1994} on one side and Gallavotti and 
Cohen~\cite{Gallavotti1995a} on the other, leading respectively to the transient
and steady-state fluctuation relations. Particular attention has been
given to the question of their universality. Three decades later, they are 
understood, for Time-Reversal Invariant (TRI) systems, as a statistical 
refinement of the second law of thermodynamics.

Note that a parallel literature concerning fluctuation relations for work
(instead of entropy) developed in the late 1990's, following the seminal
contributions of Jarzynski and Crook~\cite{Jarzynski1997a,Crooks1998}.
In this context, a first version of such relations was stated in the
earlier works by Bochkov and Kuzovlev~\cite{Bochkov1977,Bochkov1979,Bochkov1981b}, 
and their contribution was then rediscovered. See the recent 
monograph~\cite{Cuneo2026} and references therein for more details.

The transient (Evans--Searles) and the steady (Gallavotti--Cohen) fluctuation
theorems are two mathematically and physically distinct, although deeply
connected, results. Both theorems involve a large deviation principle for the
mean entropy production rate (in the sequel, whenever the meaning is clear, we
will often abbreviate it as \textsl{entropy production}). Recall that,  given two
classical states $\mu$, $\nu$,  \ie two probability measures on a phase space
$\cX$,  their relative entropy $\Ent(\nu|\mu)$ is the expectation of a function of their
Radon--Nikodym derivative $\frac{\d\nu}{\d\mu}$, see
Equation~\eqref{def:classicrelativeentropy}, and satisfies $\Ent(\nu|\mu)\leq0$
with equality iff $\nu=\mu$. When applied to an initial state $\omega$ and its
time evolution $\omega_t$, the so-called entropy balance equation reads
\[
\Ent(\omega_t|\omega) = -\omega(c^t)
\]
for some suitable random variable $c^t$ called entropic cocycle, and which
corresponds to the phase space contraction over the time interval $[0,t]$ (see
Equation~\eqref{def:classicalcocycle} and below). The inequality
$\omega(c^t)=-\Ent(\omega_t|\omega)\geq0$  expresses the second law of
thermodynamics in the form of non-negativity of the mean entropy produced over
the interval $[0,t]$. This is a statement about the expectation value of $c^t$;
it does not assert monotonicity in $t$ or non-negativity of the instantaneous
entropy production. Fluctuation theorems concern a more precise statistical
description of the random process $(c^t)_{t\geq 0}$, also interpreted as the
phase space contraction, and its law $ Q_{\nu, t}$ with respect to a given
measure $\nu$.

For Time-Reversal Invariant (TRI) systems, the transient fluctuation relation
corresponds to the choi\-ce $\nu=\omega$ and asserts that, for any finite time
$t$,
\beq\label{eq:introtransient}
\frac{\d\bar Q_{\omega,t}}{\d Q_{\omega,t}}(s)=\e^{-s},
\eeq
where $\bar Q_{\omega,t}(B)=Q_{\omega,t}(-B)$ is the reflected measure, a relation 
which directly reads as a refinement of the second law (see Section~\ref{sec:classic}).

The terminology \textsl{``fluctuation theorem''} refers to the large time limit, in
the Large Deviation Principle (LDP) regime, of the mean entropy production
observable $\frac{c^t}{t}$. LDP roughly means (see Definition~\ref{def:ldp} for
a precise definition) that in the large $t$ limit one has
\[
\mathrm{Prob}_\omega\left(\frac{c^t}{t}\in B\right)
=Q_{\omega,t}(tB)\simeq \e^{-t\inf_{s\in B}I(s)},
\]
for suitable intervals $B\subset\R$ and some function $I:\R\to[0,+\infty]$
called the rate function. The relation~\eqref{eq:introtransient} then induces
the corresponding symmetry
\beq\label{eq:introratefunctionsymmetry}
I(-s)=I(s)+s.
\eeq
 
We emphasize that in the above formulation, the statistics  of the random
variable $c^t$  is taken with respect to an initial, time-reversal invariant,
state $\omega $ and it pertains to a transient process. Typically, the system
is expected to reach a (non-equilibrium) stationary regime, \ie the state
$\omega$ is expected to evolve in large time towards a non-equilibrium steady
state $\omega_\infty$. In the Gallavotti--Cohen conceptual framework,  the
essence of the non-equilibrium regime is captured  by the analysis of
$Q_{\omega_\infty, t}$, \ie the statistics of the random process
$(c^t)_{t\geq 0}$  with respect to the steady state $\omega_\infty$. The finite
time symmetry, Relation~\eqref{eq:introtransient}, breaks down but
Relation~\eqref{eq:introratefunctionsymmetry} is expected to hold true for a
corresponding rate function $I_\infty$ in  the large time limit (provided an LDP
holds).
 
To derive these LDP's, the Laplace transform 
$\ds\fF_{\nu,t}(\alpha)\coloneq\int_\RR\e^{-\alpha s}\d Q_{\nu, t}(s)$ 
of the measures $Q_{\nu,t}$ (think of $\nu$ being either $\omega$ or 
$\omega_\infty$), and in particular the large $t$ limits
\beq\label{eq:introentropicfunctionals}
\bF_\nu(\alpha)=\lim_{t\to\infty}\frac{1}{t}\log\fF_{\nu,t}(\alpha),
\eeq
play a crucial role. Indeed, one common route to an LDP is via the
Gärtner--Ellis theorem (see Theorem~\ref{thm:GEapplied}) which relies on
$\bF_\nu(\alpha)$. They will be the central objects of the theory and we will
refer to the quantities $\fF_{\nu, t}(\alpha)$ with the generic name of
\textsl{entropic functionals}. Note that for $\nu=\omega$ the
symmetry~\eqref{eq:introtransient} is equivalent to
\[
\fF_{\omega, t}(\alpha)=\fF_{\omega, t}(1-\alpha), 
\]   
while, in the Gärtner--Ellis setting where the rate function and the limiting
entropic functional are Legendre duals, the rate function
symmetry~\eqref{eq:introratefunctionsymmetry} is equivalent to
$\bF_\omega(\alpha)=\bF_\omega(1-\alpha)$ on the corresponding domains.

In typical cases, the two states $\omega$ and $\omega_\infty$ are mutually
singular,  therefore the corresponding fluctuation theorems are mathematically
very distinct statements. In spite of these differences, in all known examples
where both the transient and steady fluctuation relation hold,  it turns out
that the respective LDP rate functions $I$ and $I_\infty$ are \textsl{identical}.
Under the same duality hypotheses, the equality $I=I_\infty$ amounts to an
exchange of limits related to the entropic functionals (see Equation~\eqref{cl-ex-lim-1}). This connection was explored in~\cite{Jaksic2011},  where
the authors advocate this identity should be regarded itself as a deep dynamical
property of the system, and baptized \textsl{``Principle of Regular Entropic
Fluctuations''} (PREF).

\medskip
A lot of attention has then been devoted to the extension of fluctuation
theorems to the quantum domain. This  extension has proven to be more delicate
than anticipated,  both for conceptual and technical reasons. It came as a
surprise that the natural route through standard quantization did not lead to a
theory satisfying FT. This has been first noted in the early work of Bochkov
and Kuzovlev~\cite{Bochkov1977}, and re-attempted in the 1990's where the
approach was often referred to as ``naive quantization.'' In the work related
literature this fact is often expressed as ``work is not an observable.'' 
This approach is briefly presented in Section~\ref{ssec:naive}.

A quantum version of the \textsl{transient FT} was first obtained in the early
2000's, thanks to the contributions by Kurchan and Hal
Tasaki~\cite{Kurchan2000}, \cite{Tasaki2000} on one side and by Shuichi Tasaki
and Matsui~\cite{Tasaki2003} on the other,  these two sets of work following
different lines of reasoning. From the physical point of view,
Kurchan~\cite{Kurchan2000} and Tasaki~\cite{Tasaki2000} introduced an approach
based on two-time measurement statistics. In this approach, the main object of
interest is the distribution of the random variable expressing the entropy
change \textsl{between two successive measurements}. This observation led to the first
transient  quantum fluctuation theorem in the context of finite quantum systems.
We will refer to this approach as 2-Time Measurement (2TM) and we detail it
in Section~\ref{ssec:finitettm}. From a mathematical point of view, starting
with Araki's definition of relative entropy of states on a von Neumann algebra
in terms of relative modular operator (see Section~\ref{ssec:modular}),  Tasaki
and Matsui~\cite{Tasaki2003} proposed a quantum analogue of $Q_{\omega,t}$ that
satisfies FR, see Section~\ref{ssec:tasakimatsui}. While appealing from a
mathematical perspective, this approach, which reveals a deep link with the rich
structure of modular theory, lacks an a priori physical interpretation. However,
surprisingly the above two approaches turned out to lead to identical
entropy-production distributions at each fixed time,  shedding an unexpected new
light both on quantum mechanics probabilistic rules and modular theory.  This
connection was explored in~\cite{Jaksic2010b}  where a pedagogical  presentation
is given (see also~\cite{DeRoeck2009}  for an early work on the subject).

One drawback of the proposal by Tasaki and Matsui is that the relative modular
operator, on which the approach is based, is not an observable making it difficult
to generalize to arbitrary states. One may however rewrite the Laplace
transforms of the Tasaki and Matsui measures $Q_{\omega,t}^\tm$ in terms of the 
Araki--Connes cocycle observables~\eqref{eq:ConnesDef}. This opens another route
to the  extension of fluctuation theorems to the quantum domain. We will present
this approach in Sections~\ref{ssec:finiteqpsc} and~\ref{ssec:infiniteqpsc}, and
will refer to it as the \textsl{``quantum phase space contraction''} for reasons 
which will be transparent later.

Coming back to the physical perspective, note that the quantum measurement
protocol now enters in the definition of the object of interest at a
\textsl{fundamental level} . The distributions $Q_{\omega,t}$ correspond
specifically to a two-time \textsl{direct} measurement. Such direct measurements
have then caused debate due to a lack of experimental feasibility,  raising
interest in \textsl{indirect} measurement protocols, and it is therefore natural
to study these other measurement protocols as well. In~\cite{Benoist2025a},  we
have considered the Entropic Ancilla State Tomography (EAST) approach, where the 
system of interest is coupled to an additional one called ancilla on which the
measurement is performed, and the  corresponding entropic functional
$\fF_{\nu,t}^\anc$ is introduced (here $\nu$ refers to the state of the open
system when it is coupled to the ancilla). As we shall see, when $\nu=\omega$,
this approach also leads to the same distributions $Q_{\omega,t}$ (see
Section~\ref{ssec:ancilla}).

\medskip
Turning to the \textsl{steady} version of the FT, an extension to quantum two-time
measurement entropy production was for a long time missing from the literature.
The series of work~\cite{Benoist2024a,Benoist2025b,Benoist2025a,Benoist2026}
provides such an extension and explores the quantum version of the PREF. These
works focus on open quantum systems, where a small system interacts with several
infinitely extended reservoirs, and which play a privileged role in the study
of non-equilibrium quantum statistical mechanics. The purpose of this review is
to present, in an introductory way, the approach proposed in these works.

First we underline that in order to obtain steady states with nonzero entropy
production in the Hamiltonian setting considered here, one has to consider
infinitely extended reservoirs, described either via a thermodynamic limit or
directly using an algebraic approach, while the 2TM approach is by essence
limited to reservoirs either finite dimensional or confined with discrete energy
spectra. A procedure allowing to define entropy production statistics with
respect to $\omega_\infty$, the quantum analogue of $Q_{\omega_\infty,t}$, is
provided in the algebraic setting by~\cite{Benoist2024a}. The key
formula~\eqref{def:ttmextended} allows to construct, for infinitely extended
reservoirs, finite time entropic functionals $\fF^\ttm_{\omega_T,t}(\alpha)$
associated to the time evolved states $\omega_T$. They are the Laplace transform
of Borel probability measures which are naturally identified as the
entropy-production distributions of a 2TM protocol where the first measurement
is taken at time $T$ and the second at $T+t$.  A large $T$ limit finally allows
to define the steady state entropic functionals
$\fF^\ttm_{\omega_\infty,t}(\alpha)$ and associated probability distributions.
The emerging quantum steady fluctuation theorem in 2TM however comes with a
surprising degree of rigidity due to the invasive thermodynamic effect of the
first measurement in the thermodynamic limit.  For infinitely extended systems
with ergodic reservoirs, the consequence is dramatic: the modular averaging that
represents the first measurement resets the reservoir part of a normal initial
state to the reference reservoir state, the unique invariant state in its normal
folium (see Section~\ref{ssec:infinitettm}).  As a  consequence,  the PREF
quantum analog trivializes in the 2TM approach.

The quantum phase space contraction and the indirect,  ancilla assisted,
measurement approaches are both also natural. They coincide with the 2TM
description when statistics are considered with respect to the initial state
$\omega$, while this identity need not hold for other states. It turns out that,
while the associated entropic functionals $\fF^\ttm_{\omega_\infty,t}(\alpha)$,
$\fF^\anc_{\omega_\infty,t}(\alpha)$ and $\fF^\qpsc_{\omega_\infty,t}(\alpha)$
generally differ for finite $t$, under additional strong, although natural, 
ergodic assumptions, their large $t$ behavior as given
by~\eqref{eq:introentropicfunctionals} actually coincide. Together with the
rigidity of the 2TM approach this leads to a more general version of a quantum
PREF, see Definition~\ref{def-PREF}.

We emphasize that in the indirect measurement and quantum phase space
contraction approaches, the primary objects are the functionals
$\fF^\anc_{\nu,t}$ and $\fF^\qpsc_{\nu,t}$.  In other words, these approaches
are defined at the level of ``Laplace transform''.  However, when starting from
an arbitrary state $\nu$,  the functionals $\fF_{\nu,t}^\anc$ and
$\fF^\qpsc_{\nu,t}$ are  in general not the Laplace transform of any
probability measure. Consequently there is no guarantee that these
generalizations are associated to random processes.

It can also be easily seen that, when considering the indirect measurement
approach, the quantum PREF actually relies on a suitable exchange of limits
which is typically a deep dynamical problem that has to be proven model by
model, see Section~\ref{ssec:prefeast}. This approach parallels in this sense
the classical theory. We emphasize that the indirect measurement approach relies
on a quantum state tomography (performed on the auxiliary system). One therefore
needs to repeat the indirect measurement experiment with the \textsl{same}
initial state, \ie the state of the open quantum system at the time
where it is coupled to the ancilla. In this respect the steady state is a
natural choice leading therefore to the study of the functional
$\fF^\anc_{\omega_\infty,t}(\alpha)$. The PREF in its strongest form, \ie
for the steady state, is thus crucial.

This being said, a weaker version of the PREF, with finite waiting time,  holds
in greater generality in the following sense. If the open system, initially in
the state $\omega$, has already evolved for some time $T$ before it is coupled
to the ancilla,  the large $t$ behavior, again given
by~\eqref{eq:introentropicfunctionals}, of $\fF^\anc_{\omega_T,t}(\alpha)$ does
not depend on $T$ (see Equation~\eqref{eq:finitetimeeastpref}). This reflects a
stability of the process all along the trajectory.

As mentioned above, in its indirect measurement version (but also in the phase
space contraction one) the quantum PREF is a deep, ergodic type, dynamical
problem. In~\cite{Benoist2025a},  we considered a spectral approach to it and
show how to characterize this quantum PREF in terms of spectral resonances of
quantum transfer operators.  We developed an axiomatic scheme to analyze these
resonances, building on techniques introduced
in~\cite{Jaksic1996a,Jaksic1996b,Jaksic2002a}. The application of this scheme is
carried out in~\cite{Benoist2026} for a paradigmatic model of open quantum
system,  namely the spin-fermion model.  We provide a summary in
Section~\ref{sec:quantumtransferoperator}.

This review is organized as follows. In Section~\ref{sec:classic} we briefly
recall the entropic Fluctuation Theorem and Relation for classical systems. The
quantum framework used in the rest of the paper is given in
Section~\ref{sec:quantumdescription}. In Section~\ref{sec:finitetransient} we
explain the various approaches to transient fluctuation relation in the context
of finite quantum systems. Section~\ref{sec:infinitesystems} is devoted to the
generalization of the two-time measurement, indirect measurement and quantum
phase space contraction entropic functionals for open systems with infinitely
extended reservoirs and associated to arbitrary states. The quantum PREF is
discussed in Section~\ref{sec:quantumpref} and the quantum transfer operator
method in Section~\ref{sec:quantumtransferoperator}.

\medskip
\noindent{\bf Acknowledgments.} The work of CAP and VJ was partly funded by the 
CY Initiative grant Investissements d'Avenir, grant number ANR-16-IDEX-0008. 
VJ acknowledges the support of NSERC and the support of the MUR grant 
"Dipartimento di Eccellenza 2023-2027" of Dipartimento di Matematica, Politecnico 
di Milano. We also acknowledge the support of the ANR project DYNACQUS, grant 
number ANR-24-CE40-5714.

\section{Classical systems}\label{sec:classic}

In this section we briefly recall the entropic fluctuation theorems for
classical systems in order to put in perspective the issues and results which
appear in the quantum context. This section is based on~\cite{Jaksic2011} where
the reader can find a complete exposition and proofs.

The setup is a pair $(\cX,\phi)$, where $\cX$ is a compact metric space and
$\phi=\{\phi^t\mid t\in\RR\}$ is a group of homeomorphisms of $\cX$ such that
the map
\[
\RR\times\cX\ni(t,x)\mapsto\phi^t(x)\in\cX
\]
is continuous.  We denote by $C(\cX)$ the vector space of all continuous
complex-valued functions on $\cX$ and equip it with the sup norm
$\|f\|_{\infty}\coloneq \sup_{x\in\cX}|f(x)|$.  Observables are functions $f\in
C(\cX)$, and they evolve in time as $f\mapsto f_t=f\circ\phi^t$. States are
Borel probability measures on $\cX$, and we write $\nu(f)=\int_\cX f\d\nu$. They
evolve in time as $\nu \mapsto\nu_t=\nu\circ\phi^{-t}$. A state $\nu$ is
$\phi$-invariant if $\nu_t=\nu$ for all $t$. The relative entropy of two states
$\nu$ and $\mu$ is defined by
\beq\label{def:classicrelativeentropy}
\Ent(\nu|\mu)=
\begin{cases}
-\int_\cX\log\left(\frac{\d\nu}{\d\mu}\right)\d\nu&\text{if }\nu\ll\mu;\\[4pt]
-\infty&\text{otherwise},
\end{cases}
\eeq
and satisfies  $\Ent(\nu|\mu)\leq0$ with equality iff $\nu=\mu$.

A time-reversal of $(\cX,\phi)$ is an involutive homeomorphism $\iota:\cX\to\cX$
such that $\iota\circ\phi^t=\phi^{-t}\circ\iota$ for all $t\in \RR$. Given such
a time-reversal map $\iota$, a state $\nu$ is called Time-Reversal Invariant
(TRI) if $\nu\circ\iota=\nu$.

The starting point is then a triple $(\cX,\phi,\omega)$ where, to avoid
triviality, the reference (initial) state $\omega$ is supposed not to be
$\phi$-invariant. The system is TRI if $\omega$ is TRI for some time-reversal of
$(\cX,\phi)$. If for all $t\in \RR$ the measures $\omega$ and $\omega_t$ are
mutually absolutely continuous, we consider the following entropic cocycle
\beq\label{def:classicalcocycle}
c^t\coloneq \left(\log \frac{\d\omega_t}{\d\omega}\right) \circ \phi^t
= - \log \frac{\d\omega_{-t}}{\d\omega}.
\eeq
Its naturalness comes from the following entropy balance equation 
\[
\Ent(\omega_t|\omega)=-\omega(c^t).
\]
It is also the phase space contraction between time $0$ and time $t$: for any Borel 
set $B\subset X$ one has 
\[
\int_B \e^{-c^t}\d\omega=\int_{\phi^t(B)} \d\omega.
\]
We will regard $c^t$ as a random variable over the probability space $(\cX,\omega)$ 
and denote by $Q_{\omega,t}$ its law w.r.t.\;$\omega$. The entropy balance equation 
then becomes
\beq\label{eq:ctaverage}
\int_\R s \, \d Q_{\omega,t}(s) = -\Ent(\omega_t|\omega).
\eeq
As mentioned in the introduction, the non-positivity of relative entropy
together with the entropy balance equation~\eqref{eq:ctaverage} gives the
non-negativity of the mean entropy produced over $[0,t]$.

The family $(c^t)_{t\in\RR}$ is an additive $\phi$-cocycle, \ie
$c^{t+s}=c^t+c^s\circ\phi^t$ for all $t,s\in\RR$, and if we moreover assume that
the map $\RR\ni t\mapsto c^t\in C(\cX)$ is differentiable at $t=0$ the {\sl
	entropy production} observable (or {\sl phase space contraction rate}) of
$(\cX,\phi,\omega)$ is defined by
\beq\label{def:classicalentropyprod}
\sigma=\frac{\d\ }{\d t}c^t\big|_{t=0}.
\eeq
One then has $c^t=\int_0^t \sigma \circ \phi^s \, \d s$ so that 
$\Ent(\omega_t|\omega)=-\int_0^t \omega_s(\sigma)\d s$.

The cocycle $(c^t)_{t\in\RR}$ enjoys the following additional properties:
\begin{itemize}
\item[1.] Because $\int_\cX \e^{-c^t}\, \d \omega =1$, either $c^t=0$
$\omega$-almost surely or $c^t$ takes both positive and negative values with
positive $\omega$-probability.
\item[2.] If $(\cX,\phi,\omega)$ is TRI we further have, for all $t\in \RR$,
\[
c^t\circ \iota = c^{-t} =-c^t\circ \phi^{-t},
\]
so that the image of $c^t$ is actually symmetric around $0$.
\item[3.] The inequality $ \int_\R s\,\d Q_{\omega,t}(s)\geq0$  shows that, with
respect to $\omega$, positive values of \(c^t\) are favored on average, even
though \(c^t\) may take both positive and negative values (whether or not the
system is TRI).
\end{itemize}

\medskip
Entropic Fluctuation Theorem and Relation go beyond these simple observations.
When the system is TRI one actually gets that for all $t\in\RR$
\beq
\frac{\d\bar Q_{\omega,t}}{\d Q_{\omega,t}}(s)=\e^{-s}, 
\label{str-2-cl-1}
\eeq
where $\bar Q_{\omega,t}(B)=Q_{\omega,t}(-B)$ is the reflected measure. All
together, this can be summarized as
\begin{quote}
Whenever its law is non-trivial, $c^t$ takes both positive and negative values,
in a symmetric range for TRI systems.  Moreover, for such TRI systems and with
respect to the initial $\omega$, positive values of $c^t$ are favored in an
exponential way with a universal exponential rate.
\end{quote}

The terminology \textsl{Entropic Fluctuation Theorem} actually refers to a (full
or restricted) LDP for the family of measures $(P_{\omega,t})_{t>0}$,
$P_{\omega, t}(B)=Q_{\omega, t}(tB)$, that is the law of the variables
\[
\frac{c^t}{t} = \frac{1}{t}\int_0^t \sigma\circ\phi^s \, \d s.
\]
For the sake of completeness we recall the basic definition and refer
to~\cite{Dembo2000,Ellis2006} for more information about LDP.
\begin{definition}\label{def:ldp} The family of probability measures
$(P_t)_{t>0}$ satisfies a \textbf{full LDP} if there exists a lower-semicontinuous 
function $\II:\RR\to[0,\infty]$, called the \textbf{rate function}, such that 
for any Borel set $B\subset \RR$,
\beq
-\inf_{s\in\interior(B)}\II(s)\leq \liminf_{t\to\infty}\tfrac1t\log P_t(B) 
\leq \limsup_{t\to\infty}\tfrac1t\log P_t(B)\leq -\inf_{s\in\closure(B)}\II(s),
\label{ldp-rate-11}
\eeq
where $\interior(B)/\closure(B)$ denotes the interior/closure of $B$.

If~\eqref{ldp-rate-11} holds only for Borel sets $B\subseteq\, ]a,b[$, where
$]a, b[\not=\RR$, we then say that a \textbf{restricted LDP} holds for
$(P_t)_{t>0}$ on the interval $]a,b[$.
\end{definition}

If such an LDP holds for $(P_{\omega,t})_{t>0}$, the relation~\eqref{str-2-cl-1} 
and the induced symmetry 
$$
\II(-s)=\II(s)+s
$$
of the LDP rate function for $(P_{\omega,t})_{t>0}$, on the domain where both 
$s$ and $-s$ are covered by the LDP, constitute the so-called \textsl{classical 
Evans--Searles (or transient) fluctuation relation.}

The next step is then to go beyond the initial/reference state $\omega$. For any
state $\nu$ on $\cX$, we denote by $Q_{\nu,t}$ the law of $c^t$ w.r.t. $\nu$,
and $P_{\nu, t}(B)=Q_{\nu, t}(tB)$. Of particular interest is the case where
$\nu$ is a NESS of $(\cX,\phi,\omega)$, \ie a weak-limit point of the net
\[
\left\{\frac1T\int_0^T \omega_t\d t\,\bigg|\, T>0\right\}
\]
as $T\uparrow \infty$. The set of NESS is non-empty and any NESS is $\phi$-invariant. 
Moreover, for any NESS $\omega_\infty$ one has
\[
\omega_\infty(\sigma)\geq 0.
\]
Suppose that the weak limit $\ds\lim_{t\to\infty}\omega_t=\omega_\infty$ exists so that, 
in particular, $\omega_\infty$ is the unique NESS of the system $(\cX, \phi, \omega)$. 
If the full/restricted LDP holds for $(P_{\omega_\infty,t})_{t>0}$, we say that the
\textsl{full/restricted classical Gallavotti--Cohen fluctuation theorem} holds for 
$(\cX,\phi,\omega)$. If in addition the respective rate function $\II_\infty$ satisfies
\beq
\II_\infty(-s)= \II_\infty(s) +s
\label{classical-gc-1}
\eeq
on its domain of definition, we say that the \textsl{classical Gallavotti--Cohen fluctuation
relation} holds. Unlike in the Evans--Searles case, \eqref{classical-gc-1} is not ensured by 
the LDP and time-reversal invariance. In any model where the symmetry~\eqref{classical-gc-1} 
has been proven it turns out that $\II_\infty=\II$.

As mentioned in the introduction, one usual route to an LDP relies on the Laplace transform 
via the Gärtner--Ellis theorem.
\begin{theorem}\label{thm:GEapplied}
Let $I=]\vartheta_-,\vartheta_+[\subset\RR$ be an open interval containing $0$, and
suppose that the limit
\[
F(\alpha)=\lim_{t\to\infty}\frac{1}{t}\log \int_\RR e^{-\alpha t s}\d P_t(s)
\]
exists,  is finite for $\alpha\in I$, and that the function  $F:I\to\RR$ is differentiable. 
If $I=\RR$, then the full LDP holds with the rate function
\beq\label{ldp-rate-21}
\II(s)=\sup_{-\alpha\in I}(s\alpha-F(-\alpha)).
\eeq
Otherwise, the restricted LDP with rate function~\eqref{ldp-rate-21} holds on the interval 
$]a, b[$ where
$$
a=-\lim_{\alpha\uparrow\vartheta_+}F^\prime(\alpha),\qquad
b=-\lim_{\alpha\downarrow\vartheta_-}F^\prime(\alpha).
$$
\end{theorem}

For any state $\nu$, $t\in\RR$ and $\alpha\in \CC$, let
\begin{equation}\label{eq:classicallaplace}
\fF_{\nu,t}(\alpha)\coloneq\int_\RR \e^{-\alpha ts}\d P_{\nu, t}(s)
=\int_\RR \e^{-\alpha s}\d Q_{\nu, t}(s)
=\int_\cX\e^{-\alpha c^t}\d\nu
=\int_\cX \left(\frac{\d\omega_{-t}}{\d\omega}\right)^{\alpha}\d\nu
\end{equation}
denote the Laplace transform of $Q_{\nu,t}$. We further set
\beq\label{def:GEfunction}
\bF_{\nu}(\alpha) \coloneq  \lim_{t\to\infty}\frac{1}{t}\log \fF_{\nu, t}(\alpha),
\eeq
whenever the limit exists.

Note that the transient fluctuation relation~\eqref{str-2-cl-1} is equivalent to the symmetry
$$
\fF_{\omega,t}(\alpha)=\fF_{\omega,t}(1-\alpha), \quad \forall\alpha\in\R.
$$
When the rate functions and limiting entropic functionals are Legendre duals, as
in the Gärtner--Ellis setting above, the basic properties of the Legendre
transform show that $\II=\II_\infty$ is equivalent to
$\bF_\omega=\bF_{\omega_\infty}$ on the corresponding domains. For each fixed
$T$, the density $\d\omega_T/\d\omega$ is bounded above and bounded away from
zero, so that $\bF_{\omega_T}=\bF_\omega$ whenever $\bF_\omega$ exists.
Moreover, the weak convergence $\omega_T\to\omega_\infty$ gives
$\fF_{\omega_T,t}(\alpha)\to\fF_{\omega_\infty,t}(\alpha)$ for each fixed $t$.
Thus, the equality $\bF_\omega=\bF_{\omega_\infty}$ is equivalent to the
following exchange of limits:
\beq\label{cl-ex-lim-1}
\lim_{T\to\infty}\lim_{t\to\infty}\tfrac1t\log\fF_{\omega_T,t}(\alpha)
=
\lim_{t\to\infty}\lim_{T\to\infty}\tfrac1t\log\fF_{\omega_T,t}(\alpha).
\eeq
This exchange of limits is the only known general mechanism that ensures the validity of 
the Gal\-lavot\-ti--Cohen fluctuation theorem and relation. The identities
\beq\label{eq:classicalpref}
\bF_\omega=\bF_{\omega_\infty}, \qquad \II=\II_\infty
\eeq
have been raised to the classical \textsl{Principle of Regular Entropic Fluctuations} (PREF), 
see~\cite{Jaksic2011}.

\section{Quantum framework}
\label{sec:quantumdescription}

\subsection{$C^\ast$ quantum dynamical systems}
\label{ssec:cstar}

A $C^\ast\!${\sl-quantum dynamical system} is a triple $(\cO,\tau,\omega)$ where:
\begin{itemize}
\item $\cO$ is a $C^\ast$-algebra with a unit $\one$. Its elements $A\in\cO$ describe the
system observables.
\item $\RR\ni t\mapsto\tau^t$ is a $C^\ast$-dynamics, \ie a strongly continuous
group of $\ast$-automorphisms of $\cO$. It  describes the Heisenberg
time-evolution $A_t=\tau^t(A)$ of the system observables. The infinitesimal
generator of a $C^\ast$-dynamics $\tau$  is a possibly unbounded
$\ast$-derivation $\delta$ of $\cO$. We use the convention $\tau^t=\e^{t\delta}$.
\item $\omega$ is a state on $\cO$, \ie a continuous linear functional such that 
$\omega(A^\ast A)\ge0$ for all $A\in\cO$, and $\omega(\one)=1$. The number $\omega(A)$ 
is the quantum expectation value of the observable $A\in\cO$, when the system is in the 
state $\omega$. States evolve according to the Schrödinger picture $\omega_t=\omega\circ\tau^t$, 
so that $\omega_t(A)=\omega(A_t)$. $\omega$ is called $\tau$-invariant whenever $\omega_t=\omega$ 
for all $t\in\RR$.
\end{itemize}

Thermal equilibrium of the system at inverse temperature $\beta\in\RR^\ast$ is described by a 
state $\omega$ satisfying the $(\tau,\beta)$-KMS boundary condition: for any $A,B\in\cO$ the 
function $F_{A,B}(t)=\omega(A\tau^t(B))$ has an analytic extension to the complex strip 
$\{z\mid 0<\mathrm{sign}(\beta)\Im z<|\beta|\}$,
which is bounded and continuous on its closure, and satisfies
$$
F_{A,B}(t+\i\beta)=\omega(\tau^t(B)A)
$$
for all $t\in\RR$. Such states are said to be $(\tau,\beta)$-KMS, and are $\tau$-invariant.
The quantum dynamical system $(\cO,\tau,\omega)$ is said to be
in thermal equilibrium at inverse temperature $\beta\in\RR$ if its reference state $\omega$ 
is $(\tau,\beta)$-KMS. 

For the modular framework below, we assume that the GNS representation
associated to $\omega$ is faithful; this is automatic for the CAR and spin
algebras considered here. A state $\omega$ is called modular if it is
$(\varsigma_\omega,-1)$-KMS for some $C^\ast$-dynamics $\varsigma_\omega$. When
it exists $\varsigma_\omega$ is unique and is called the modular group of
$\omega$. We denote its generator by $\delta_\omega$. If $\omega$ is
$(\tau,\beta)$-KMS then it is modular and $\varsigma_\omega^t=\tau^{-\beta t}$.
This modular group will play a crucial role in the description of the two-time
measurement approach to quantum Entropic Fluctuations, see
Section~\ref{ssec:infinitettm}.

The quantum dynamical system $(\cO,\tau,\omega)$ is called ergodic if $\omega$ is
 $\tau$-invariant and 
\[
\lim_{T\to\infty}\frac{1}{T}\int_0^T\omega\left(B^\ast\tau^t(A)B\right)\,\d t
=\omega(A)\omega(B^\ast B)
\]
holds for all $A,B\in\cO$. 

Finally, as we have seen in the classical case, time-reversal plays an important role. 
In the quantum domain, a time-reversal of $(\cO,\tau)$ is an anti-linear
involutive $\ast$-automorphism $\Theta$ of $\cO$ satisfying
\[
\Theta\circ\tau^t=\tau^{-t}\circ\Theta
\]
for all $t\in\RR$. A state $\omega$ is TRI if there exists a time-reversal $\Theta$ 
such that $\omega(\Theta(A))=\omega(A^\ast)$ for all $A\in\cO$.

\medskip
\textbf{Example 1: finite systems.} For quantum systems with finite dimensional
Hilbert space $\cK$, $\cO$ is the $C^\ast$-algebra of all linear maps on $\cK$,
states are identified with density matrices, the dynamics is given by
$\tau^t(A)=\e^{\i tH}A\e^{-\i tH}$ where $H$ is the Hamiltonian of the system,
and thermal equilibrium states are given by the usual Gibbs states $\rho_\beta=
\e^{-\beta H}/\Tr\left(\e^{-\beta H}\right)$.  Finite systems with $\dim\cK>1$
and a faithful reference state are never ergodic.

\textbf{Example 2: ideal Fermi gas,} with single particle Hilbert space $\fh$ 
and single particle Hamiltonian $h$. The algebra of observables is
the CAR-algebra $\cO=\CAR(\fh)$, the $C^\ast$-algebra generated by 
creation/annihilation operators $a^\ast(f)$/$a(f)$, $f\in\fh$, 
satisfying the canonical anti-commutation relations
$$
\{a(f),a^\ast(g)\}=\langle f,g\rangle\one,\qquad
\{a(f),a(g)\}=0.
$$
The dynamics on $\cO$ is the group of Bogoliubov $\ast$-auto\-morphisms
associated to $h$, \ie the $C^\ast$-dyna\-mics defined by
$\tau^t(a(f))=a\left(\e^{\i th}f\right)$. The unique $(\tau,\beta)$-KMS state on
$\cO$ is the gauge-invariant quasi-free state $\omega$ whose two-point function
is given by $\ds \omega(a^\ast(f)a(g)) = \bra g, (1+\e^{\beta h})^{-1} f\ket$. If
$h$ has purely absolutely continuous spectrum then $(\cO,\tau,\omega)$ is
ergodic (it is actually mixing).

\subsection{Open quantum systems}\label{ssec:oqs}

In these notes, we will focus on a special class of quantum dynamical systems,
the so-called open quantum systems, which play a privileged role in the study of
non-equilibrium quantum statistical mechanics.  They consist of a small system
$\sS$ interacting with several reservoirs.  For simplicity of exposition, we
will consider only the case of $2$ reservoirs and refer the reader to the
series~\cite{Benoist2024a,Benoist2025a,Benoist2025b,Benoist2026} for the general
situation.

As mentioned in the introduction, we will need to consider infinitely extended
reservoirs.  In the literature they are described either as a thermodynamic
limit of finite dimensional systems, or directly using the algebraic setting
described in the previous section.  In this contribution,  we shall use both
descriptions. The finite dimensional approximations are needed to introduce the
two-time measurement approach while the more general setting  is needed to
consider steady states.

In the algebraic general setting, each thermal reservoir $\sR_j$, $j=1,2$, is
described by a $C^\ast$-quantum dynamical system $(\cO_j,\tau_j,\omega_j)$,
$\tau_j^t=\e^{t\delta_j}$. We assume them to be initially in thermal equilibrium
at some inverse temperature $\beta_j>0$, that is we assume that $\omega_j$ is a
$(\tau_j,\beta_j)$-KMS state. These reservoirs are coupled through a finite
dimensional system $\sS$ described by $(\cO_\sS,\tau_\sS,\omega_\sS)$ where
$\cO_\sS=\cB(\cK_\sS)$ and  $\tau_\sS$ is generated by the Hamiltonian $H_\sS$.
For convenience, we assume that $\omega_\sS$ is the tracial state. Analogous
results hold for any faithful $\tau_\sS$-invariant state.

In the absence of interaction, the combined system $\sS+\sR_1+\sR_2$ is
described by the quantum dynamical system $(\cO,\tau_\mathrm{fr},\omega)$,
where\footnote{Whenever the meaning is clear within the context, we write $A$
for $A\otimes\one$ and $\one\otimes A$, $\delta_j$ for
$\delta_j\otimes\mathrm{Id}$, $\mathrm{Id}\otimes\delta_j$, etc.}
$$
\cO=\cO_\sS\otimes \cO_1\otimes\cO_2, \qquad
\tau_\mathrm{fr}=\tau_\sS\otimes \tau_1\otimes\tau_2,\qquad 
\omega=\omega_\sS\otimes \omega_1\otimes\omega_2.
$$
We write $\omega_\sR=\omega_1\otimes\omega_2$ for the product reservoir state.
The state $\omega$ is modular with 
\beq\label{eq:modulargroupoqs}
\varsigma_\omega^\theta
=\mathrm{Id}_\sS\otimes\tau_1^{-\beta_1 \theta}\otimes \tau_2^{-\beta_2 \theta}. 
\eeq

The interaction of\/ $\sS$ with $\sR_j$ is described by a self-adjoint element 
$V_j\in\cO_\sS\otimes\cO_j$, and  the full interaction by $V=V_1+V_2$. The 
interacting dynamics $\tau$ is generated by 
\[
\delta=\delta_\mathrm{fr}+\i[V,\argdot],\qquad  
\delta_\mathrm{fr}=\i[H_\sS,\argdot]+\delta_1+\delta_2. 
\]
For later purposes, see Section~\ref{ssec:modular}, we shall also assume that 
\assuming{Reg}{Reg}{%
$V_j\in \rm{Dom}(\delta_j)$, $j=1,2$.
}

When the reservoirs $\sR_j$ are also finite dimensional systems, with non-interacting 
Hamiltonians $H_{\sR_j}$, this means that the evolution of the coupled system is 
generated by the Hamiltonian 
\beq\label{eq:finiteham}
H=H_\mathrm{fr}+V,\qquad 
H_\mathrm{fr}=H_\sS+H_{\sR_1}+H_{\sR_2},
\eeq
and the initial state given by the density matrix
\beq\label{eq:finiteomega}
\omega = \frac{\mathrm{Id}_\sS}{{\rm dim}\, \cK_\sS} 
\otimes \frac{\e^{-\beta_1 H_{\sR_1}}}{\tr \left(\e^{-\beta_1 H_{\sR_1}}\right)}
\otimes \frac{\e^{-\beta_2 H_{\sR_2}}}{\tr \left(\e^{-\beta_2 H_{\sR_2}}\right)}. 
\eeq

When considering the Gallavotti--Cohen fluctuation theorem, we will also make 
the following assumption
\assuming{NESS}{NESS}{%
The system $(\cO,\tau,\omega)$ has a unique NESS $\omega_\infty$, and for all $A\in\cO$ one has
$$
\lim_{t\to\infty}\omega\circ\tau^t(A)=\omega_\infty(A).
$$
}

\subsection{Relative entropy and entropy observable in finite systems}

For finite systems, the relative entropy of a pair of states $\mu,\nu$
\footnote{Recall that for finite systems states are identified with density matrices}
is given by
\begin{equation}\label{eq:RelEntDefFinite}
\Ent(\nu|\mu)=
\begin{cases}
\tr\left( \nu \log \mu -\nu\log \nu\right) &\text{if }\nu\ll\mu;\\[4pt]
-\infty&\text{otherwise}.
\end{cases}
\end{equation}
where $\nu\ll\mu$ means $\ker \mu \subset \ker \nu$. The following entropy balance 
equation 
\[
\Ent(\omega_t|\omega) = -\int_0^t \omega_s(\sigma)\, \d s
\]
relates the relative entropy of the evolved state $\omega_t$ with respect to 
$\omega$ to the entropy production observable 
\beq\label{def:epobservablefinite}
\sigma = -\beta_1\Phi_1 - \beta_2\Phi_2, \qquad \Phi_j=\i[H_{\sR_j},V_j].
\eeq

For general systems and an arbitrary pair of states, relative entropy has been 
introduced by Araki and requires modular structure, see Equation~\eqref{eq:RelEntDef} 
in the next section.

\subsection{Modular structure}\label{ssec:modular}

In this section, we provide a brief account of the modular structure associated
to a quantum dynamical system. This structure has played a fundamental role in
the modern development of quantum statistical mechanics, including the entropic
fluctuations theorems presented in this review. We shall only describe the main
objects, mainly to fix notation. For a more complete introduction we refer the
reader to the literature, see e.g. any of the references~\cite{Bratteli1981,
Bratteli1987,Derezinski2003a,Haag1996,Ohya1993,Stratila1981}. For the unfamiliar 
reader's convenience, the modular structure of finite systems will be described 
in the next section.

We denote by $(\cH,\pi,\Omega)$ the GNS representation of $\cO$ associated to
the state $\omega$, by $\fM=\pi(\cO)''$ its enveloping von Neumann
algebra.\footnote{$\fA'$ denotes the commutant of a subset $\fA\subset\cB(\cH)$,
	\ie the set of elements of $\cB(\cH)$ which commute with all elements of $\fA$.}
The state $\omega$ has a unique normal extension to $\fM$ which we denote by the
same symbol. A density matrix $\rho$ on $\cH$ defines a state on $\fM$ by the
familiar quantum mechanical rule $\fM\ni A\mapsto\tr(\rho A)$. Such states on
$\fM$ are called {\sl normal,} and their restriction to $\pi(\cO)$ induce states
on $\cO$ which are called $\omega$-{\sl normal.} The {\sl folium} of $\omega$ is
the set $\cN$ of all $\omega$-normal states on $\cO$ (it is the quantum analogue
of the set of measures absolutely continuous with respect to a given measure
$\omega$). We assume in this modular framework that $\omega$ is modular and that
$\tau$ extends to a pointwise $\sigma$-weakly continuous group of
$\ast$-automorphisms of $\fM$. These regularity properties hold for the bounded
open-system perturbations considered here.

Since the state $\omega$ is modular (see Section~\ref{ssec:cstar}), its normal
extension to $\fM$ is faithful, \ie the map $\fM\ni A\mapsto A\Omega\in\cH$ is
injective. In this case, it follows from Tomita--Takesaki's theory that the GNS
Hilbert space $\cH$ comes with a modular structure:
\begin{itemize}
\item The anti-linear operator defined on the dense subspace
$\pi(\cO)\Omega\subset\cH$ by $\pi(A)\Omega\mapsto\pi(A)^\ast\Omega$ has a
closure $S$ with the unique polar decomposition $S=J\Delta^{1/2}_\omega$. There,
$\Delta_\omega$ is a densely defined positive linear operator and $J$ an
anti-unitary involution on $\cH$. $\Delta_\omega$ is called \textsl{modular
operator} and $J$ \textsl{modular conjugation} of the pair $(\fM,\omega)$. The
closure $\cH_+$ of the subset $\{ AJA\Omega\mid A\in\fM \}$ is its \textsl{natural
	cone}. The modular operator implements the modular group on $\pi(\cO)$, \ie
$$
\pi\big(\varsigma_\omega^\theta(A)\big)
=\Delta_\omega^{\i\theta} \pi(A) \Delta_\omega^{-\i\theta},
$$
and thus allows to extend this group to $\fM$.

\item Any normal state $\nu$ on $\fM$ has a unique vector representative
$\Omega_\nu\in\cH_+$ such that, for all $A\in\fM$, 
$\nu(A)=\langle\Omega_\nu,A\Omega_\nu\rangle$.
Moreover, a unit vector $\Psi\in\cH_+$ is cyclic for $\fM$,
\ie $\fM\Psi$ is dense in $\cH$, iff the state it induces on $\fM$ is faithful,
\ie the map $\fM\ni A\mapsto A\Psi\in\cH$ is injective.

\item If $\mu$ and $\nu$ are faithful normal states on $\fM$, the densely
defined anti-linear map $A\Omega_\nu\mapsto A^\ast\Omega_\mu$ has a closure
$S_{\mu|\nu}$ whose  polar decomposition reads
$S_{\mu|\nu}=J\Delta^{1/2}_{\mu|\nu}$, $J$ being the modular conjugation of
$(\fM,\omega)$ and $\Delta_{\mu|\nu}$ a densely defined positive linear
operator: the \textsl{relative modular operator} of the pair $(\mu,\nu)$.

\item Let $(\mu,\nu)$ be a pair of faithful normal states on $\fM$. Their
relative entropy is defined by
\beq
\Ent(\nu|\mu)=\langle\Omega_\nu,\log\Delta_{\mu|\nu}\Omega_\nu\rangle.
\label{eq:RelEntDef}
\eeq
This is the original definition by Araki~\cite{Araki1975/76,Araki1977} with the
sign and ordering convention of~\cite{Jaksic2001a}.

\item The Araki--Connes cocycle of this pair is the strongly continuous 
one-parameter family of unitaries in $\fM$ given by\footnote{Here, we depart from
the traditional convention, denoting by $[D\mu:D\nu]_{\i t}$ what is usually
written $[D\mu:D\nu]_t$.}
\beq
[D\mu:D\nu]_\alpha=\Delta_{\mu|\nu}^\alpha\Delta_\nu^{-\alpha},\qquad(\alpha\in\i\RR).
\label{eq:ConnesDef}
\eeq
We mention that, contrary to the unitaries $\Delta_{\mu|\nu}^\alpha$ and
$\Delta_\nu^{-\alpha}$, the Araki--Connes cocycle belongs to $\fM$, \ie is an
observable. This cocycle, and in particular the quantities
$[D\omega_{-t}:D\omega]_\alpha$, will be central to the theory, see
Section~\ref{sec:infinitesystems}. Assumption \Reg{} ensures that the latter
actually belong to $\pi(\cO)$, see e.g.~\cite[Lemma 2.4]{Benoist2025b}. Denoting
$\pi^{-1}([D\omega_{-t}:D\omega]_{\alpha})\in \cO$ by
$[D\omega_{-t}:D\omega]_{\alpha}$ this allows to consider quantities like
$\nu\left([D\omega_{-t}:D\omega]_{\alpha}\right)$ for arbitrary states on $\cO$,
not necessary $\omega$-normal, in particular the NESS $\omega_\infty$.

\item  The standard Liouvillean $\cL$ is the unique self-adjoint operator on the
GNS space $\cH$ satisfying
$$
\e^{-\i t\cL}\cH_+=\cH_+,\qquad \tau^t(A)=\e^{\i t\cL}A\e^{-\i t\cL},
$$
for all $t\in\RR$ and $A\in\fM$. In particular, $\Omega_t\coloneq \e^{-\i t\cL}\Omega$ 
is the vector representative of $\omega_t=\omega\circ\tau^t$ in the
natural cone. When $\omega$ is $(\tau,\beta)$-KMS (this is the case for the
reservoir Liouvilleans $\cL_{\sR_j}$), then $\cL=-\frac{1}{\beta}
\log\Delta_\omega$. Let us just mention that, see {\sl e.g.}~\cite{Jaksic2012},
for any two faithful normal states $\mu$ and $\nu$ one has
\beq
\e^{-\i t\cL}\Delta_{\nu|\mu}\e^{\i t\cL}=\Delta_{\nu\circ\tau^t|\mu\circ\tau^t}.
\label{eq:DeltaCovar}
\eeq Finally, we note that the ergodic properties of a quantum dynamical
system $(\cO,\tau,\omega)$ are related to the spectral properties of $\cL$.  For
a $\tau$-invariant reference state $\omega$, $(\cO,\tau,\omega)$ is ergodic if
and only if $0$ is a simple eigenvalue of $\cL$.

For the open systems considered in Section~\ref{ssec:oqs} the standard
Liouvillean of the interacting system has the form
\[
\cL = \cL_\free + V -JVJ, \quad \cL_\free=\cL_\sS+\cL_{\sR_1}+\cL_{\sR_2}.
\]
\end{itemize}

\subsection{Modular structure of finite systems}
\label{ssec:finitemodular}

For the  unfamiliar reader's convenience,  let us give the explicit form of a
GNS representation and of the modular data for quantum systems on a finite
dimensional Hilbert space $\cK$ and initial state given by the positive definite
density matrix $\omega$.  In this case $\cH=\cB(\cK)$ endowed with the scalar
product $\bra X,Y\ket = \tr(X^\ast Y)$, $\pi(A)$ is the left multiplication by $A$,
$\pi(A):X\mapsto AX$ and $\Omega=\omega^{1/2}$. The modular operator and
conjugation are
\[
\Delta_\omega:X\mapsto \omega X \omega^{-1}, \qquad J:X\mapsto X^\ast,
\] 
and the modular group is given by
\[
\varsigma_\omega^t(A)=\omega^{it}A\omega^{-it}.
\] 
The relative modular operator of a pair $(\mu,\nu)$ is 
\[
\Delta_{\mu|\nu}:X\mapsto \mu X \nu^{-1},
\] 
and~\eqref{eq:RelEntDef} indeed coincides with~\eqref{eq:RelEntDefFinite}. Finally 
the Araki--Connes cocycle is 
\[
[D\mu:D\nu]_\alpha:X\mapsto \mu^{\alpha}\nu^{-\alpha}X
\] 
which indeed belongs to $\pi(\cO)$. Of particular importance in the next
sections will be the observable
$[D\omega_{-t}:D\omega]_\alpha=\omega_{-t}^\alpha\omega^{-\alpha}$. Finally the
standard Liouvillean associated to the dynamics generated by the Hamiltonian $H$
is $\cL=[H,\argdot]$.

\noindent\textbf{Remark 1.} As already mentioned the Araki--Connes cocycles
$[D\mu:D\nu]_\alpha$ will play a fundamental role. For finite systems they are
well defined and belong to  $\pi(\cO)$ for all $\alpha\in\C$. In the general
case it is initially defined only for  $\alpha\in\i\RR$,  and additional
hypothesis will be needed to guarantee that they belong to $\pi(\cO)$. The
latter property is needed to consider their expectation value with respect to
the (non $\omega$-normal) steady state  $\omega_\infty$.

\noindent\textbf{Remark 2.} Classical systems also fit into the algebraic framework, the
algebra of observables being $C(\cX)$. In that case the modular operator
associated to a state $\omega$ is trivial (it is the identity), the relative
modular operator of a pair $(\mu,\nu)$ is related to their Radon--Nikodym
derivative, namely $\Delta_{\mu|\nu}: g\mapsto \frac{\d\mu}{\d\nu}g$ while the
Connes cocycle is given by
\begin{equation}\label{eq:classicalcocycle}
[D\mu:D\nu]_\alpha = \left(\frac{\d\mu}{\d \nu}\right)^\alpha.
\end{equation}

\section{Finite systems and the transient fluctuation theorem}
\label{sec:finitetransient}

Although infinitely extended systems are required to consider situations which
are thermodynamically non-trivial, in this section we restrict ourselves to
finite dimensional systems in order to make the definition of the various
entropic functionals under consideration more transparent (this is particularly
important for the two-time measurement approach). Throughout this section, we
therefore assume that the reservoirs, and consequently the entire system, are
finite dimensional. We denote by $H_{\sR_j}$ the non-interacting Hamiltonian of
$\sR_j$.

\subsection{Naive quantization}
\label{ssec:naive}

The entropic cocycle
\begin{equation}\label{def:cocyclefinite}
c^t  = -\log\omega_{-t}+\log\omega,
\end{equation}
is the natural quantization of its classical counterpart,
see~\eqref{def:classicalcocycle}, and the entropy production observable $\sigma$
defined in~\eqref{def:epobservablefinite} satisfies (compare
with~\eqref{def:classicalentropyprod})
\beq\label{eq:epobservablefinite}
\sigma=\left.\frac{\d c_t}{\d t}\right|_{t=0}.
\eeq

It is thus natural to consider, as the quantum analogue of the measure
$Q_{\nu,t}$ of Section~\ref{sec:classic}, the spectral measure 
$Q^\mathrm{naive}_{\nu,t}$ of $c^t$ with respect to the state $\nu$ whose 
Laplace transform is thus given by
\begin{equation}\label{eq:naivelaplace}
\fF^\mathrm{naive}_{\nu,t}(\alpha)=\nu \left( \e^{-\alpha c^t}\right)
=\nu \left( \e^{\alpha \log\omega_{-t}-\alpha\log\omega}\right) .
\end{equation}

The basic properties of $c^t$ hold at the quantum level:
\begin{enumerate}
\item It follows from $\tr(c^t)=0$ that, unless $c^t=0$, its spectrum contains both 
positive and negative values,
\item One still has $0\geq\Ent(\omega_t|\omega)=-\omega(c^t)$ so that, with respect 
to \(\omega\), positive values of \(c^t\) are favored on average,
\item If the system is TRI with time-reversal $\Theta$ then $\Theta(c^t)=-\tau^{-t}(c^t)$ 
so that the spectrum of $c^t$ is symmetric around $0$.
\end{enumerate}
However, the transient fluctuation relation fails. If $\bar Q^\mathrm{naive}_{\omega,t}$ 
is the reflected measure associated to $Q^\mathrm{naive}_{\omega,t}$, one gets, using 
TRI, that for any function $f$
\[
\int_\R f(s) \, \d \bar Q^\mathrm{naive}_{\omega,t} 
=\omega\left( f(c^t) \e^{\log \omega_{-t}}\e^{-\log \omega}\right)
\]
while
\[
\int_\R f(s) \e^{-s}\, \d Q^\mathrm{naive}_{\omega,t}
=\omega \left( f(c^t) \e^{\log\omega_{-t}-\log\omega}\right).
\]
At the level of the Laplace transform this is reflected by the fact that
\begin{equation}\label{eq:naivelaplaceomega}
\fF^\mathrm{naive}_{\omega,t}(\alpha)
=\omega \left( \e^{\alpha \log\omega_{-t}-\alpha\log\omega}\right)
\end{equation}
does not satisfy the symmetry 
$\fF^\mathrm{naive}_{\omega,t}(\alpha)=\fF^\mathrm{naive}_{\omega,t}(1-\alpha)$.

From a mathematical perspective, the non-commutativity of $\omega_{-t}$ and $\omega$ 
is at the origin of the failure, in general, of the fluctuation relation:
\[
\frac{\d\bar Q^\mathrm{naive}_{\omega,t}}{\d Q^\mathrm{naive}_{\omega,t}}(s)\neq\e^{-s}.
\]
Physically, the naive approach describes a single measurement of $c^t$, whereas the 
two-time measurement approach describes the entropy change between two measurements.

This failure motivated the early
searches~\cite{Kurchan2000,Tasaki2000,Tasaki2003} for alternative candidates for
quantum Evans--Searles Fluctuation Theorem and Relation. 
In~\cite{Kurchan2000,Tasaki2000} the physical aspects are considered,  and their
proposal is based  on a line of reasoning  that takes into account the
measurement issue.  The proposal ~\cite{Tasaki2003} is  mathematically oriented
and,  in a sense,  resolves the non-commutativity issue.  We shall discuss the
two approaches respectively in Sections~\ref{ssec:finitettm}
and~\ref{ssec:tasakimatsui}. Surprisingly these two proposals turned out to be
equivalent, shedding an unexpected light on both quantum mechanical
probabilistic rules and modular theory. A pedagogical discussion  of this topic
can be found in the lecture notes~\cite{Jaksic2010b}.

\subsection{Two-time measurement (2TM) entropy production}
\label{ssec:finitettm}

This notion of entropy production was introduced
in~\cite{Kurchan2000,Tasaki2000}. It consists in two consecutive measurements of
the entropy observable $S=-\log\omega$ associated to the reference state
$\omega$ given by~\eqref{eq:finiteomega}. For the open systems considered here,
\beq\label{eq:Sobservable}
S=-\log\omega = \beta_1 H_{\sR_1}+\beta_2 H_{\sR_2}+\mathrm{const}.
\eeq
For the reference state $\omega$, the same entropy-change statistics are
obtained by measuring the energies of both reservoirs at the beginning and end
of the evolution, with
$$
\Delta S=\beta_1\Delta E_1+\beta_2\Delta E_2,
$$
where $\Delta E_j$ is the measured change in energy of the $j^\mathrm{th}$ reservoir.

Note that the entropic cocycle can then be written as $c^t= \tau^t(S)-S$, the
entropy production observable as $\sigma=\frac{\d }{\d t}\tau^t(S)\lceil_{t=0}$
and that
\[
\Ent(\omega_t|\omega) = -\omega\left( \tau^t(S)-S\right).
\]

The two-time measurement protocol goes as follows. The system is in a state
$\nu$ at time $t_i$ where a first measurement of $S$ is performed. The outcome
$s_i\in \sp(S)$ is observed with probability $ p_\nu(s_i)=\nu(P_{s_i})$ where
$P_{s_i}$ is the eigenprojection corresponding to the eigenvalue $s_i\in\sp(S)$. 
After the measurement the system is in the reduced state
$$
\frac{1}{p_\nu(s_i)}P_{s_i}\nu P_{s_i},
$$
which evolves under the system dynamics over the time interval of length
$t=t_f-t_i$. The second measurement, performed at time $t_f$, yields the outcome
$s_f\in\sp(S)$ with probability
$$
p_{\nu,t}(s_f|s_i)=\frac{1}{p_\nu(s_i)}\tr\left(\e^{-\i tH}P_{s_i}\nu P_{s_i}\e^{\i tH}P_{s_f}\right).
$$
Finally, the probability of observing the pair $(s_f,s_i)$ in this  two-time
measurement protocol is
\beq
p_{\nu,t}(s_f,s_i)=p_{\nu,t}(s_f|s_i)p_\nu(s_i)
=\tr\left(\e^{-\i tH}P_{s_i}\nu P_{s_i}\e^{\i tH} P_{s_f}\right).
\label{ss-n}
\eeq
The entropy production random variable $\cE:\sp(S)\times\sp(S)\to\R$ is defined by
$$
\cE(s_f,s_i)=s_f-s_i,
$$
and its probability distribution with respect to $p_{\nu,t}$ is denoted by $Q_{\nu,t}^\ttm$,
\beq\label{def:ttmprobadistrib}
Q^\ttm_{\nu,t}(s)=\sum_{s_f-s_i=s}p_{\nu,t}(s_f,s_i).
\eeq

The family $(Q^\ttm_{\nu, t})_{t>0}$ thus describes the statistics of the two-time measurement 
entropy production. It follows from~\eqref{ss-n}-\eqref{def:ttmprobadistrib} and the functional 
calculus that their Laplace transforms are given by
\begin{equation}\label{def:ttmfinite}
\fF^\ttm_{\nu,t}(\alpha) \ = \  \int_\R\e^{-\alpha s}\d Q^\ttm_{\nu,t}(s) \ = \ \bar\nu(\omega_{-t}^\alpha\omega^{-\alpha}), \qquad (\alpha\in\C), 
\end{equation}
where $\bar\nu=\sum_{s\in\sp(S)}P_s\nu P_s$ is the diagonal part of $\nu$ with
respect to $S$.\footnote{Since $S=-\log\omega$, $\bar \nu$ is equivalently the
diagonal part of $\nu$ with respect to $\omega$.} The fact that one replaces the
initial state $\nu$ by its diagonal part $\bar\nu$ reflects the decoherence
effect induced by the first measurement.

When $\nu=\omega$, we simply have
\beq\label{def:ttmfiniteES}
\fF^\ttm_{\omega,t}(\alpha)=\omega(\omega_{-t}^\alpha\omega^{-\alpha})
=\tr(\omega_{-t}^\alpha\omega^{1-\alpha}),
\eeq
compare with~\eqref{eq:naivelaplace}. When the system is TRI one easily infers that $\fF^\ttm_{\omega,t}(\alpha)=\fF^\ttm_{\omega,t}(1-\alpha)$ for any $\alpha\in\RR$ 
restoring the transient fluctuation relation
\begin{equation}\label{eq:transientfr}
\frac{\d\bar Q^\ttm_{\omega,t}}{\d Q^\ttm_{\omega,t}}(s)=\e^{-s}. 
\end{equation}
Note also that the two distributions $Q^\ttm_{\omega,t}$ and $Q^\mathrm{naive}_{\omega,t}$ 
actually have the same first and second moment. In particular one still has 
\[
\Ent(\omega_t|\omega) = - \int_\R s\, \d Q^\ttm_{\omega,t}(s).
\]

\subsection{Tasaki--Matsui relative modular approach}
\label{ssec:tasakimatsui}

Motivated by Araki's definition of relative entropy 
\[
\Ent(\omega_t|\omega) = \bra \Omega_t | \log \Delta_{\omega|\omega_t} \Omega_t \ket = \bra \Omega | \log \Delta_{\omega_{-t}|\omega} \Omega \ket,
\]
(the second identity follows from~\eqref{eq:DeltaCovar}), in~\cite{Tasaki2003}
the authors considered the operator $-\log\Delta_{\omega_{-t}|\omega}$ as the
entropy production observable between time $0$ and time $t$, although the latter
does not belong to the algebra of observables. If $Q^\tm_{\omega,t}$ denotes the
associated spectral measures with respect to the vector $\Omega$, its Laplace
transform is therefore
$$
\fF^\tm_{\omega,t}(\alpha) = \bra \Omega,  \Delta_{\omega_{-t}|\omega}^\alpha \Omega \ket.
$$
For finite systems the relative modular operator acts as, see the end of 
Section~\ref{sec:quantumdescription},
\[
\Delta_{\omega_{-t}|\omega}: X\mapsto \omega_{-t} X \omega^{-1}.
\]
Hence (recall $\Omega=\omega^{1/2}$)
\begin{equation}\label{eq:finitetmlaplace}
\fF^\tm_{\omega,t}(\alpha) = \tr(\omega_{-t}^\alpha\omega^{1-\alpha}) 
= \omega(\omega_{-t}^\alpha\omega^{-\alpha}),
\end{equation}
which, for TRI systems, satisfies the symmetry 
$\fF^\tm_{\omega,t}(\alpha)=\fF^\tm_{\omega,t}(1-\alpha)$. The family of spectral 
measures $(Q^\tm_{\omega,t})_t$ therefore satisfies the transient fluctuation relation.

Note also that
$$
-\log\Delta_{\omega_{-t}|\omega}:X\mapsto -\log(\omega_{-t})X+X \log\omega,
$$
so the entropy production ``observable'' $-\log\Delta_{\omega_{-t}|\omega}$ is
actually the sum of two \textsl{commuting} maps, the left multiplication by
$-\log\omega_{-t}$ and the right multiplication by $\log\omega$. Although it was
not mentioned in~\cite{Tasaki2003}, the choice of
$-\log\Delta_{\omega_{-t}|\omega}$ can thus be seen as a mathematical trick to
restore the ``lack of commutativity'' mentioned in Section~\ref{ssec:naive}.
Indeed, see Section~\ref{ssec:finitemodular}, $-\log\Delta_{\omega_{-t}|\omega}$
differs from the entropic cocycle observable
\[
\pi(c^t) : X\mapsto (-\log(\omega_{-t})+ \log\omega)X
\] 
by 
\beq\label{eq:logmodular2}
\log \Delta_\omega : X\mapsto \log(\omega)\,X-X\log(\omega).
\eeq
Namely,
\beq\label{eq:relativevscocycle}
-\log\Delta_{\omega_{-t}|\omega} = \pi(c^t) -\log \Delta_\omega.
\eeq
Since $\bra \Omega, \log\Delta_\omega \Omega \ket=0$ , we get that
\[
 -\omega(c^t)=-\bra\Omega, \pi(c^t)\Omega\ket 
 = \bra \Omega | \log \Delta_{\omega_{-t}|\omega} \Omega \ket
\]
are two equivalent ways to express the relative entropy $\Ent(\omega_t|\omega)$.
The difference between the two quantities $\pi(c^t)$ and 
$-\log\Delta_{\omega_{-t}|\omega}$ however reappears at the level of the Laplace
transforms and is at the origin of the difference 
between~\eqref{eq:naivelaplaceomega} and~\eqref{eq:finitetmlaplace}. The
non-triviality of the modular operator $\Delta_\omega$ is the crucial difference
between the classical (commutative) and the quantum (non-commutative) case.

One potential objection to this approach could  be the lack of physical
interpretation. In particular, as previously mentioned,
$\log\Delta_{\omega_{-t}|\omega}$ does not even belong to the algebra of
observables.  However, its unexpected connection with the two-time measurement
process resolves this issue. Indeed, as the reader may have noticed, the
quantities $\fF^\tm_{\omega,t}(\alpha)$ and $\fF^\ttm_{\omega,t}(\alpha)$
coincide.  Therefore,  the spectral measure $Q^\tm_{\omega,t}$ is nothing but the
probability distribution $Q^\ttm_{\omega,t}$ obtained via the two-time
measurement protocol giving  an observational status to the spectral measure of
the relative modular operator. This deep link between the statistics of the
two-time measurement entropy production and modular theory has a somewhat
unusual history and  was discussed in detail in~\cite{Jaksic2010b}.

\subsection{Quantum phase space contraction (QPSC)}
\label{ssec:finiteqpsc}

The spectral projections of \(\log\Delta_{\omega_{-t}|\omega}\) do not, in
general, belong to the algebra of observables. The associated spectral measures
can therefore be defined only for states which are $\omega$ normal, leaving the
steady state $\omega_\infty$ typically out of reach. One can however further
rewrite the functional $\fF^\tm_{\omega,t}$ in terms of the Araki--Connes
cocycle, the latter having the advantage to be an observable. Namely, since
$\Delta_\omega\Omega=\Omega$ one may write
\[
\fF^\tm_{\omega,t}(\alpha) 
= \bra \Omega,  \Delta_{\omega_{-t}|\omega}^\alpha \Delta_\omega^{-\alpha}\Omega \ket  
=  \bra \Omega,  [D\omega_{-t}:D\omega]_\alpha \Omega \ket 
= \omega([D\omega_{-t}:D\omega]_\alpha).
\]
Now this allows to define an analogous family for any state $\nu$, namely
\begin{equation}\label{def:qpscfinite}
\fF^\qpsc_{\nu,t}(\alpha) \coloneq  \nu([D\omega_{-t}:D\omega]_\alpha) 
= \nu(\omega_{-t}^\alpha\omega^{-\alpha}),
\end{equation}
which we will call the \textsl{Quantum Phase Space Contraction} (QPSC) functional. 

\noindent\textbf{Remark 1.} In the classical setting, $\fF_{\nu,t}(\alpha)$ is the
Laplace transform of the classical phase space contraction $c^t$.
Combining~\eqref{eq:classicallaplace} with~\eqref{eq:classicalcocycle}
one sees that it can be expressed as
$$
\fF_{\nu,t}(\alpha)=\nu([D\omega_{-t}:D\omega]_\alpha).
$$
This makes the quantity~\eqref{def:qpscfinite} a natural quantum analogue 
and explains the terminology QPSC.

\noindent {\bf Remark 2.} When $\nu=\omega$ the QPSC functional
$\fF^\qpsc_{\omega,t}$ coincides with the Tasaki--Matsui one
$\fF^\tm_{\omega,t}$, hence with the two-time measurement functional
$\fF^\ttm_{\omega,t}$. However, for $\nu\neq \omega$ the equality with the
two-time measurement functional $\fF_{\nu,t}^\ttm$ is usually broken unless
$\nu=\bar \nu$ (compare with~\eqref{def:ttmfinite}), \ie when $\nu$ is
completely decoherent with respect to $\omega$.

\medskip
Finally, let us emphasize that, contrary to $\fF_{\nu,t}^\ttm$,
$\fF^\qpsc_{\nu,t}$ is not necessarily the Laplace transform of a Borel
probability measure when $\nu\neq\omega$ (or more generally when $\nu\neq \bar
\nu$). Thus, this family does not in general define entropy-production
probability distributions.

\subsection{Entropic Ancilla State Tomography (EAST)}
\label{ssec:finiteancilla}

As mentioned in the introduction, the two-time measurement protocol described in
Section~\ref{ssec:finitettm} has caused some debate due to a lack of
experimental feasibility, in particular for large quantum systems. This
motivated the search for more operational approaches, raising interest in
\textsl{indirect} measurement protocols. In the closely related literature
concerning work,  these points are subject to an ongoing  discussion aiming to
provide an  unambiguous definition of work,  see e.g.~\cite{DeChiara2018} and
references therein, and  a successful protocol has been introduced.  The idea is
to couple the system to an additional ancilla in order to implement physically
the Fourier transform.  A tomography on the ancilla's part then provides access
to the Fourier transform of the desired distribution. To understand this
protocol let us first consider the case where one would like to have access to
the distribution $P_{A,\nu}$ of a given measured observable $A$ with respect to
a state $\nu$. Its Fourier transform is then given by
\[
\fF_{A,\nu}(\alpha) = \int_\R \e^{-s\alpha} \d P_{A,\nu}(s) 
= \nu\left( \e^{-\alpha A}\right), \qquad \alpha=\i u\in \i\RR.
\]
The ancilla coupling procedure goes as follows:
\begin{enumerate}
\item Add a qubit and initialize it to the state $\rho_a=\frac12\begin{bmatrix}1&1\\1&1\end{bmatrix}$.
\item Let the system and qubit interact for a time $u=-\i\alpha$ with interaction Hamiltonian 
$\frac12A\otimes\sigma_z$. The qubit is then in the state 
\[
\rho_a(\alpha)=\hbox{$\tr_\cH\left(\e^{-\frac{\alpha}2A\otimes\sigma_z}\nu\otimes\rho_a
\e^{\frac{\alpha}2A\otimes\sigma_z}\right)$} 
=\frac12
\begin{bmatrix}
1&\fF_{A,\nu}(\alpha)\\
\overline{\fF_{A,\nu}(\alpha)}&1
\end{bmatrix}.
\]
\item A quantum state tomography of the qubit (using the $\sigma_x$ and $\sigma_y$ observables) 
gives then access to the Fourier transform $\fF_{A,\nu}(\alpha)$.
\end{enumerate}

In the two-time measurement protocol, one does not measure the observable
$S=-\log \omega$ once but twice, the system having evolved in the meantime. We
then consider the distribution of the differences between the two measured
values (the first value has thus to  be taken into account with a minus sign). 
The corresponding ancilla coupling procedure is similar to the previously
described one, replacing the evolution $\e^{-\frac{\alpha}2A\otimes\sigma_z}$ by
$$
\e^{-\frac{\alpha}2S\otimes\sigma_z}\e^{-\i tH\otimes\one}
\e^{+\frac{\alpha}2S\otimes\sigma_z}, \qquad S=-\log\omega.
$$

The experimental feasibility of such an evolution is again a priori not obvious. However, 
for open systems, using~\eqref{eq:finiteham} and~\eqref{eq:Sobservable} one simply has
\beq\label{eq:ancillahamilt}
\e^{-\frac{\alpha}{2} S\otimes \sigma_z}\e^{-\i t H\otimes \one}
\e^{\frac{\alpha}{2} S \otimes \sigma_z} = \e^{-\i t\widehat H_\alpha}, 
\qquad \widehat H_\alpha = H \otimes \one + \widehat W_\alpha,
\eeq
where
\beq\label{eq:ancillaperturb}
\widehat W_\alpha=\tfrac12W_\alpha\otimes(\one+\sigma_z)
+\tfrac12W_{-\alpha}\otimes(\one-\sigma_z), 
\qquad W_\alpha=\varsigma_\omega^{-\i\alpha/2}(V)-V.
\eeq
In a $2\times 2$ block matrix form this corresponds to
\[
\widehat H_\alpha=\begin{bmatrix}H+W_\alpha&0\\0&H+W_{-\alpha}\end{bmatrix}
=\begin{bmatrix}H_\mathrm{fr}+\varsigma_\omega^{-\i\alpha/2}(V)&0\\
0&H_\mathrm{fr}+\varsigma_\omega^{\i\alpha/2}(V)\end{bmatrix}.
\]
Only the interaction term $V$ is modified, to $\varsigma_\omega^{\pm\i\alpha/2}(V)$ 
depending on the qubit state. 

A simple calculation shows that the qubit is then in the state
\beq\label{eq:ancillastate}
\rho_{a,t}(\alpha)=\frac12\begin{bmatrix}1&\fF_{\nu,t}^\anc(\alpha)\\
\overline{\fF_{\nu,t}^\anc(\alpha)}&1\end{bmatrix},
\eeq
with
\begin{equation}\label{def:ancillafinite}
\fF_{\nu,t}^\anc(\alpha)
=\nu\left( \omega^{-\alpha/2}\omega_{-t}^{\alpha}\omega^{-\alpha/2}\right),
\end{equation}
the latter being again accessible using a quantum state tomography of the qubit. 
When the initial state of the system is $\omega$ we simply have
\[
\fF_{\omega,t}^\anc(\alpha)= \tr\left(\omega_{-t}^{\alpha}\omega^{1-\alpha}\right), 
\]
which coincides with the Fourier transform of the two-time measurement
probability distribution $Q^\ttm_{\omega,t}$, see
Equation~\eqref{def:ttmfiniteES}, giving the announced more operational approach
than the two-time measurement.

Note however that when $\nu\neq\omega$ the ancilla functional
$\fF_{\nu,t}^\anc$ generally differs from both the 2TM and the QPSC functionals
(compare~\eqref{def:ancillafinite} with~\eqref{def:ttmfinite}
and~\eqref{def:qpscfinite}). Moreover we stress that, as it is the case for the
QPSC functional, the ancilla functional $\fF^\anc_{\nu,t}$ is not necessarily
the Laplace transform of a Borel probability measure when $\nu\neq\omega$.

\section{Infinitely extended systems: 3 facets of entropic fluctuations}
\label{sec:infinitesystems}

Besides their own interest, infinitely extended systems are needed
to consider entropic fluctuations with respect to steady states. In this section
we explain how the 2TM, QPSC and ancilla approaches presented in the previous
section for finite systems can be generalized to infinitely extended systems.
The QPSC and ancilla approaches are rather straightforward to extend, although
not completely obvious. The 2TM approach is more problematic. The two-time
measurement entropy production protocol is always introduced in the context of
finite quantum systems (or, slightly more generally, confined quantum systems
with possibly infinite but discrete energy spectra), and its definition for
infinitely extended systems obtained via a thermodynamic limit (TDL). For a long
time such a limiting procedure was unknown when the state $\nu$ at the instant
of the first measurement is arbitrary, in particular for the steady state
$\nu=\omega_\infty$. To the best of our knowledge, the case of a general state
$\nu$ was considered for the first time in~\cite{Benoist2024a}.

For the sake of simplicity, and motivated by the Gallavotti--Cohen fluctuation
theorem, in the rest of this review we will only consider the case where the
state $\nu$ is either the time-evolved state $\omega_T$, for some $T\geq 0$, or
the NESS $\omega_\infty$. The conceptual and technical difference between these
two states is that,  while $\omega_T$ is $\omega$-normal for all $T\ge0$,
the NESS $\omega_\infty$ and $\omega$ are typically mutually singular. Most of
the results presented below hold for arbitrary state $\nu$, with the distinction
$T$ finite or infinite being replaced by $\nu$ is $\omega$-normal or arbitrary.

In the sequel the functions $\fF_{\omega_T,t}^\#$, $\#\in\{\ttm, \qpsc, \anc\}$ and
$T\in \R_+\cup\{\infty\}$, will be called the \textsl{entropic functionals}. As
for classical systems, they are the central objects to study LDP through the use
of the Gärtner--Ellis theorem.

\subsection{2TM protocol in infinitely extended systems}
\label{ssec:infinitettm}

\subsubsection{Definition and stability}

For finite systems the entropic functional $\fF_{\omega_T,t}^\ttm$ is given
by~\eqref{def:ttmfinite}. In terms of modular data\footnote{Modular theory is
the natural language to describe infinitely extended systems where quantities
like $-\log\omega$ do not have a meaning.} it becomes
\[
\fF_{\omega_T,t}^\ttm(\alpha)=\overline\omega_T\left([D\omega_{-t}:D\omega]_\alpha\right),
\]
where we recall that $\overline\omega_T$ is the completely decoherent part of
$\omega_T$ with respect to $\omega$.

When $T=0$ one simply has $\fF_{\omega,t}^\ttm(\alpha) = \omega
\left([D\omega_{-t}:D\omega]_{\alpha}\right)$ and the generalization to extended
systems is straightforward. When $T\neq 0$ the meaning of $\overline\omega_T$ is
not clear anymore and the key observation is that one can further write,
see~\cite{Benoist2024a},
\begin{equation}\label{def:ttmextended}
\fF_{\omega_T,t}^\ttm(\alpha) = \lim_{R\to\infty}\frac1R\int_0^R\omega_T\circ\varsigma_\omega^\theta \left([D\omega_{-t}:D\omega]_{\alpha}\right)\d\theta.
\end{equation}
The ergodic average over the modular group $(\varsigma_\omega^\theta)_\theta$,
see~\eqref{eq:modulargroupoqs}, reflects the effect of a first measurement and
replaces the projection to the diagonal part with respect to $\omega$.

\noindent\textbf{Remark.} Since the initial state $\omega$ is invariant under the
action of the modular group $\varsigma_\omega$, for $T=0$ one gets as expected
that
\beq\label{eq:ttmextendedES}
\fF_{\omega,t}^\ttm(\alpha) = \omega \left([D\omega_{-t}:D\omega]_{\alpha}\right),
\eeq
which can also be rewritten as
\beq\label{eq:ttmextendedES2}
\fF_{\omega,t}^\ttm(\alpha)
=\omega \left([D\omega_{-t}:D\omega]^\ast_{\frac{\bar\alpha}{2}}
[D\omega_{-t}:D\omega]_{\frac{\alpha}{2}}\right).
\eeq
The latter expression will be useful in comparison with the ancilla assisted
protocol, see Section~\ref{ssec:ancilla}.

In~\cite[Theorem 1.3]{Benoist2024a} it is proven that, in the modular setting
recalled in Section~\ref{sec:quantumdescription}, under the regularity
hypotheses stated there, for any $T\in\RR$, $t\in\RR$ and $\alpha\in\i\RR$ the
limit in the right-hand side of~\eqref{def:ttmextended} exists. Moreover there
is a unique Borel probability measure $Q_{\omega_T,t}^\ttm$ on $\RR$ such that
$$
\fF_{\omega_T,t}^\ttm(\alpha)=\int_\RR\e^{-\alpha s}\d Q_{\omega_T,t}^\ttm(s).
$$
The measure $Q_{\omega_T,t}^\ttm$ gives the statistics of an idealized two-time
measurement of the entropy production in the system $(\cO,\tau,\omega)$ in a
protocol where the first measurement is done at time $T$ and the second at time
$T+t$.

This definition of such an idealized two-time measurement entropy production
actually  comes with a somewhat surprising rigidity result: under mild ergodic
assumptions on the reservoirs,  it is essentially independent of $T$. More
precisely, the following is proven in~\cite[Theorem 1.6]{Benoist2024a}.
\begin{theorem}\label{thm:stability} Suppose that each reservoir subsystem is
ergodic. For all $T$ denote by $\omega_{T,\sS}$ the restriction of the evolved
state $\omega_T$ to $\cO_\sS$, and by $\Omega_{T,\sS}$ its representative
vector. We write $\Omega_\sR=\Omega_1\otimes\Omega_2$ for the representative
vector of $\omega_\sR$. Then for all $\alpha\in\i\R$,
\beq\label{eq:ttmopensyst}
\fF^\ttm_{\omega_T,t}(\alpha) = \omega_{T,\sS}\otimes\omega_\sR\left([D{\omega_{-t}}:D{\omega}]_\alpha\right) = \langle\Omega_{T,\sS}\otimes\Omega_\sR,\Delta_{\omega_{-t}|\omega}^\alpha\, \Omega_{T,\sS}\otimes\Omega_\sR\rangle.
\eeq
In particular, $Q^\ttm_{\omega_T,t}$ is the spectral measure of
$-\log\Delta_{\omega_{-t}|\omega}$ for the vector
$\Omega_{T,\sS}\otimes\Omega_\sR$. Moreover, for any $t\in \R$,
\beq\label{eq:ttmstability}
\min \sp(\omega_{T,\sS})\times \dim\,\cK_\sS\leq\frac{\d Q^\ttm_{\omega_T,t}}{\d Q^\ttm_{\omega,t}} \leq \dim\,\cK_\sS,
\eeq
where $\sp(\omega_{T,\sS})$ denotes the spectrum of the density matrix 
associated to the state $\omega_{T,\sS}$.
\end{theorem}

\begin{corollary} Under the same hypotheses, if in addition Assumption \NESS{}
holds then Equation~\eqref{eq:ttmopensyst} allows to define $\ds
\fF^\ttm_{\omega_\infty,t}=\lim_{T\to \infty} \fF^\ttm_{\omega_T,t}$. The
associated measure $Q_{\omega_\infty,t}^\ttm$ is obtained as the weak limit of
$Q_{\omega_T,t}^\ttm$ as $T\to\infty$ and~\eqref{eq:ttmstability} remains in
force.
\end{corollary}

\noindent{\bf Remark 1.} If one allows a direct interaction
$V\in\cO_1\otimes\cO_2$ between the reservoirs and takes $\dim\,\cK_\sS=1$, one
actually has the even stronger rigidity property that
$Q_{\omega_T,t}^\ttm=Q^\ttm_{\omega,t}$ for all $T$ and $t$.

\noindent{\bf Remark 2.} The needed ergodicity assumption is actually the one of
the modular groups of the reservoir states, \ie  the
$(\cO_j,\omega_j,\varsigma_{\omega_j})$'s need to be
ergodic~\cite{Benoist2024a}. This is equivalent to the above mentioned
assumption because the $\omega_j$'s are $(\tau_j,\beta_j)$-KMS.

\noindent {\bf Remark 3.} The first inequality in~\eqref{eq:ttmstability} is
non-trivial only when $\omega_{T,\sS}$ is faithful. This is automatic when $T$
is finite.

The two-time measurement entropy production statistics exhibits a striking
rigidity, as shown by Inequality~\eqref{eq:ttmstability}, which can be
interpreted as the memory erasing effect due to the decoherence induced by the
first measurement on the reservoirs. For infinitely extended systems this effect
is dramatic as the modular averaging associated with the first measurement
resets the reservoir part of a normal initial state to the reference reservoir
state, which is the unique invariant state in its normal folium, while the state
of the small system  $\sS$ has  a marginal effect, being finite dimensional. In
Section~\ref{sec:quantumpref} we shall explore its consequences concerning the
related LDP's.

\noindent{\bf Remark 4.} As mentioned, this rigidity result is due to the
ergodicity of the modular groups (of the reservoir initial states). It has no
classical analog\footnote{The modular groups are then trivial and hence not
	ergodic, except for the one-dimensional algebra.}, see~\cite[Section
2.8]{Benoist2025a} for a discussion of this point.

We conclude this section with the following proposition when $T=0$.  It shows
that in the two-time measurement approach the entropic functional can be
extended to real values of $\alpha$,  leading to a  transient fluctuation
relation  for infinitely extended systems.
\bep\label{prop:ttmtransientinfinite} The map
$\i\RR\ni\alpha\mapsto\fF_{\omega,t}^\ttm(\alpha)$ has an analytic extension to
the vertical strip $0<\Re\alpha<1$ which is bounded and continuous on its
closure.

If moreover the system is time-reversal invariant then,  for any $0\leq
\alpha\leq 1$,
$$
\fF_{\omega, t}^\ttm(\alpha)= \fF_{\omega, t}^\ttm(1- \alpha).
$$
As a consequence, the measures $Q_{\omega,t}^\ttm$ satisfy the transient 
fluctuation relation~\eqref{eq:transientfr}.
\eep

\subsubsection{Thermodynamic justification of 2TM}
\label{ssec:thermolimit}

In order to define the two-time measurement protocol for infinitely extended
systems we have used the following customary route: a physical notion,
introduced in the context of finite quantum systems, is expressed in a modular
form (here it is~\eqref{def:ttmextended}) and through this form is directly
extended, by definition, to a general quantum dynamical system. In concrete
physical models this definition is justified by the TDL limit. The latter has
been extensively studied in early days of quantum statistical mechanics and the
wealth of obtained results make its implementation in modern literature most
often a routine exercise which is generally skipped. However the mixture of
quantum measurements and thermodynamics, which is central to the definition of
two-time quantum measurement entropy production, has a number of unexpected
features among which is the above-mentioned drastic decoherence effect due to
the first measurement. For this reason the TDL justification of this idealized
two-time measurement needs to be put on solid physical grounds.

In more mathematical terms the first measurement decoherence corresponds to a
projection onto $\Ker\log\Delta_\omega$, see Equation~\eqref{eq:ttmprojection}
below. Indeed, assume that for some $B_T\in \fM'$ with $\|B_T\Omega\|=1$, one
has $\Omega_T=B_T\Omega$. This holds true, see~\cite[Proposition~2.1]{Benoist2025b}, 
if the map
$$
\i\RR\ni z\mapsto[D\omega_T:D\omega]_z
$$
has an analytic continuation to the vertical strip $0<\Re z<\frac12$ that is 
continuous and bounded on its closure, taking
\beq\label{eq:BT}
B_T=J [D\omega_T:D\omega]_{\frac12}J\in\fM'.
\eeq
One can then write~\eqref{def:ttmextended} as, using that 
$\varsigma^\theta\left([D\omega_{-t}:D\omega]_{\alpha}\right) \in \fM$ commutes 
with $B_T$,
\begin{eqnarray}
\fF_{\omega_T,t}^\ttm(\alpha)&=&\lim_{R\to\infty}\frac1R\int_0^R\langle B_T^\ast B_T\Omega,
\Delta_\omega^{\i\theta}[D\omega_{-t}:D\omega]_\alpha\Delta_\omega^{-\i\theta}\Omega\rangle\,\d\theta\nonumber\\
&=&\lim_{R\to\infty}\frac1R\int_0^R\langle B_T^\ast B_T \Omega, 
\Delta_\omega^{\i\theta}[D\omega_{-t}:D\omega]_\alpha\Omega\rangle\d\theta\nonumber\\[10pt]
&=&\langle B_T^\ast B_T \Omega, \one_{\{0\}}(\log \Delta_\omega) [D\omega_{-t}:D\omega]_\alpha \Omega \rangle, \label{eq:ttmprojection}
\end{eqnarray}
where $\one_{\{0\}}(\log \Delta_\omega)$ is the mentioned projection on 
$\Ker\log\Delta_\omega$.

For infinitely extended reservoirs, and due to the ergodicity assumptions 
(see the end of Section~\ref{ssec:modular}), this projection is given by
$$
\one_{\{0\}}(\log \Delta_\omega)=\mathrm{Id}_{\sS} \otimes |\Omega_{R_1}\ket\bra\Omega_{R_1}| \otimes  |\Omega_{R_2}\ket\bra\Omega_{R_2}|
$$
and its rank equals $\left(\mathrm{dim}\cK_\sS\right)^2$. On the contrary, 
for finite dimensional reservoirs and using~\eqref{eq:finiteomega}
and~\eqref{eq:logmodular2}, one has
\[
\log\Delta_\omega=-\beta_{\sR_1}[H_{\sR_1},\argdot]-\beta_{\sR_2}[H_{\sR_2},\argdot],
\]
so the rank of this projection is at least 
$\left( \mathrm{dim}\cK_\sS\right)^2\times\mathrm{dim}\cK_{\sR_1}\times\mathrm{dim}\cK_{\sR_2}$.
Along a sequence of TDL approximations the size of the kernel
$\Ker\log\Delta_\omega$ thus grows to infinity together with the dimension of
the reservoirs Hilbert spaces to suddenly drastically shrink in the limiting
state. This dimension reduction is at the core of the need for the TDL
justification of the definition~\eqref{def:ttmextended} of the 2TM.

\noindent\textbf{Remark.} The stability result given in
Theorem~\ref{thm:stability} relies on the representation of
$\fF_{\omega_T,t}^\ttm(\alpha)$ given in~\eqref{eq:ttmprojection}.

\medskip
If, for each reservoir $\sR_j$,
$(\cO_{j,\Lambda},\tau_{j,\Lambda},\omega_{j,\Lambda})_\Lambda$ is a net of
finite dimensional approximations of the infinitely extended $C^\ast$ quantum
dynamical system $(\cO_j,\tau_j,\omega_j)$, one thus needs to prove
that\footnote{A quantity with an index $\Lambda$ means that it is the quantity
obtained in the corresponding finite dimensional approximation.}
\[
\lim_\Lambda \langle B_{T,\Lambda}^\ast B_{T,\Lambda} \Omega_\Lambda, \one_{\{0\}}(\log \Delta_{\omega,\Lambda}) [D\omega_{-t,\Lambda}:D\omega_\Lambda]_\alpha \Omega_\Lambda \rangle = \langle B_T^\ast B_T \Omega, \one_{\{0\}}(\log \Delta_\omega) [D\omega_{-t}:D\omega]_\alpha \Omega \rangle.
\]
In~\cite{Benoist2025b} a general scheme is proposed, which can then be used
on concrete models.  We briefly comment on two specific aspects of this TDL
problem and refer the reader to~\cite{Benoist2025b} for more details.

One of the issues is the control of the projections $\one_{\{0\}}(\log
\Delta_{\omega,\Lambda})$. One proves that they converge weakly to
$\one_{\{0\}}(\log \Delta_{\omega})$ by combining Araki's results on the
continuity of modular structure~\cite[Section 2]{Araki1974b} with the dynamical
ergodicity assumption on the infinitely extended reservoirs. In the models
considered, the latter ensures the inclusion
$\Ker\log\Delta_\omega\subset\Ker\log\Delta_{\omega,\Lambda}$ needed in the
projection convergence argument. We emphasize that such an ergodicity assumption
is unusual in TDL type results. The second main issue concerns the control of
the Araki--Connes cocycles $[D\omega_{-t,\Lambda}:D\omega_\Lambda]_\alpha$, not
only for $\alpha\in\i\RR$ but also, in view of~\eqref{eq:BT}, of their analytic
extensions to a vertical strip up to $\Re(\alpha)=\frac{1}{2}$.

The general TDL scheme mentioned above is applied in~\cite{Benoist2025b} to
extended black-box fermionic models and quantum spin systems, under the
hypotheses specified there. For quantum spin systems these include interaction
decay, high temperature and reservoir ergodicity. The finite-volume reservoir
states are restrictions of the infinite-volume KMS states, with effective
Hamiltonians $\widehat
H_{j,\Lambda}=-\beta_j^{-1}\log(\omega_j\lceil\cO_{j,\Lambda})$. The
general-state TDL for the conventional finite-volume Gibbs approximation remains
open. Even with the restriction approximation, one does not prove that
\[
\lim_\Lambda \fF_{\omega_{T,\Lambda},t}^\ttm(\alpha) =\fF_{\omega_T,t}^\ttm(\alpha), \qquad T,t\in\R,\ \alpha\in\i\RR,
\]
and consequently  $\lim_\Lambda Q^\ttm_{\omega_{T,\Lambda},t} =
Q^\ttm_{\omega_T,t}$ weakly, for the net of states $\omega_{T,\Lambda} =
\omega_\Lambda\circ \tau_\Lambda^T$, \ie  the evolution of the restriction
$\omega_\Lambda$ of $\omega$ to the finite volume $\Lambda$ under the finite
volume evolution $\tau_\Lambda$. Instead,  one gets
\[
\lim_\Lambda \fF_{\omega_T\lceil\Lambda,t}^\ttm(\alpha) =\fF_{\omega_T,t}^\ttm(\alpha), \qquad T,t\in\R,\ \alpha\in\i\RR,
\]
where $\omega_T\lceil\Lambda$ is the restriction to the finite volume $\Lambda$
of the infinite volume evolved state $\omega_T$. Except in trivial cases 
$\omega_T\lceil\Lambda \neq \omega_{T,\Lambda}$.

\subsection{Entropic cocycle and QPSC}
\label{ssec:infiniteqpsc}

The second approach to fluctuation relation, the QPSC approach presented in Section~\ref{ssec:finiteqpsc}, is more straightforward to extend to infinitely extended systems than the 2TM approach. Indeed, compare with~\eqref{def:qpscfinite}, we consider the map
$$
\i \RR \ni \alpha \mapsto [D\omega_{-t}: D\omega]_{\alpha}\in \cO
$$
as a characterization of the {\em quantum phase space contraction} of $(\cO,\tau,\omega)$ at time $t$, and we set 
\beq\label{qpscDef}
\fF_{\omega_T,t}^\qpsc(\alpha)\coloneq \omega_T([D\omega_{-t}:D\omega]_{\alpha}), \qquad T\in \R\cup\{\infty\}.
\eeq
Note that when $T=0$ this functional coincides with the one obtained from the 2TM protocol, see Equation~\eqref{eq:ttmextendedES},
$$
\fF_{\omega,t}^\qpsc=\fF_{\omega,t}^\ttm.
$$
This identity need not hold when $T\neq 0$.

To motivate further the above definition and the relevance of~\eqref{qpscDef},
we shall relate it to the entropic cocycle $c^t$. As already pointed out, the
definition~\eqref{def:cocyclefinite} of $c^t$ is meaningless in infinitely
extended systems. Instead it can be  defined through the use of the
Araki--Connes family $([D\omega_t:D\omega]_\alpha)_{t\in\RR,\alpha\in\i\RR}$, 
which can be regarded as a non-commutative version of the Radon--Nikodym
derivative $\frac{\d\omega_t}{\d\omega}$.

The regularity Assumption~\Reg{} guarantees that the map
$\RR\ni\theta\mapsto[D\omega_t:D\omega]_{\i\theta}\in\cO$ is differentiable at
$\theta=0$ for all $t\in\RR$, see~\cite[Lemma 2.4]{Benoist2025b}. The entropic
cocycle $c^t$ is then defined by
\beq\label{eq:ctDef}
c^t\coloneq\left.\i\frac{\d\ }{\d\theta}[D\omega_{-t}:D\omega]_{\i\theta}\right|_{\theta=0}.
\eeq
Note that $c^t$ is a self-adjoint element of $\cO$ and indeed coincides
with~\eqref{def:cocyclefinite} for finite systems, compare also
with~\eqref{def:classicalcocycle} in the classical case.  The map 
$\RR\ni t\mapsto c^t\in\cO$ is then differentiable at $t=0$ and the entropy production
observable $\sigma$ defined by, compare with~\eqref{def:classicalentropyprod} in
the classical setting and with~\eqref{def:epobservablefinite}
and~~\eqref{eq:epobservablefinite} for finite quantum systems,
\[
\sigma\coloneq\left.\frac{\d c_t}{\d t}\right|_{t=0} = -\beta_1\delta_1(V_1)-\beta_2\delta_2(V_2).
\]
The properties of $(c^t)_t$ and $\sigma$ then completely parallel those observed
in classical systems, see Section~\ref{ssec:naive}. Namely, we have:
\begin{enumerate}
\item the family $(c^t)_{t\in\RR}$ is an additive $\tau$-cocycle, 
\ie $c^{t+s}=c^t+\tau^t(c^s)$ for all $t,s\in\RR$,
\item $-\log \Delta_{\omega_{-t}|\omega} = c^t-\log \Delta_\omega$ (compare 
with~\eqref{eq:relativevscocycle}) so that $\Ent(\omega_t|\omega)=-\omega(c^t)$.
\item $c^t=\int_0^t \sigma_s \d s$ so that 
$\Ent(\omega_t|\omega)=-\int_0^t \omega_s(\sigma)\d s$.
\item If $(\cO, \tau, \omega)$ is TRI with time-reversal 
$\Theta$, then $\Theta(c^t)=c^{-t}=-\tau^{-t}(c^t)$.
\end{enumerate}

\subsection{Entropic Ancilla State Tomography (EAST)}
\label{ssec:ancilla}

The EAST functional presented in Section~\ref{ssec:finiteancilla} can similarly
be generalized to infinitely extended systems where it becomes even more
relevant. Indeed, for such extended systems, the two-time measurement approach
raises both the question of experimental feasibility and that of the invasive
decoherence due to the first measurement, with its induced rigidity on the
probability distributions $Q_{\omega_T,t}^\ttm$.

The procedure described for finite systems can be easily generalized to
infinitely extended systems, the ancilla remaining a qubit.  The algebra of
observables of the system coupled to a qubit is $\widehat\cO\coloneq
\cO\otimes{\mathrm{Mat}_2(\CC)}$, where $\mathrm{Mat}_2(\CC)$ is the algebra of
complex $2\times2$ matrices. The interacting system+ancilla dynamics on
$\widehat\cO$ is given by, compare with~\eqref{eq:ancillahamilt},
\[
\widehat\tau_\alpha^t=\e^{t\hat\delta_\alpha},\qquad \widehat\delta_\alpha=\delta\otimes\mathrm{Id}+\i[\widehat{W}_\alpha,\argdot],
\]
where $\widehat W_\alpha$ is the perturbation defined
by~\eqref{eq:ancillaperturb} (note that it is self-adjoint). The initial state
of the ancilla is $\rho_a$ as above. For any state $\nu$ on $\cO$ we set
$\widehat\nu=\nu\otimes\rho_a$,
$\widehat\nu_t(\alpha)=\widehat\nu\circ\widehat\tau_\alpha^t$, and
$\rho_{a,t}(\alpha)=\widehat\nu_t|_{{\rm Mat}_2(\CC)}$. We then have:

\bep\label{prop:ancillafunctional}
Suppose that $V\in\Dom(\delta_\omega)$. Then, for all $\alpha \in \i\RR$ and any
state $\nu$ on $\cO$, the ancilla's state $\rho_{a,t}(\alpha)$ at time $t\in\RR$
is given by the right-hand side of~\eqref{eq:ancillastate} with
\beq\label{def:ancillainfinite}
\fF_{\nu,t}^\anc(\alpha) = \nu\left([D\omega_{-t}:D\omega]^\ast_{\frac{\bar \alpha}{2}} [D\omega_{-t}:D\omega]_{\frac{\alpha}{2}}\right).
\eeq
\eep
Here again, when $\nu=\omega$ we get
$\fF_{\omega,t}^\anc(\alpha)=\fF_{\omega,t}^\ttm(\alpha)$ (see
Equation~\eqref{eq:ttmextendedES2}), and the ancilla coupling protocol indeed
gives the announced more operational definition of entropy production.   We will
consider this ancilla protocol starting with $\nu=\omega_T$,
$T\in\R_+\cup\{\infty\}$. For $T\neq 0$ it generally differs from both the 2TM
functional~\eqref{def:ttmextended} and the QPSC one~\eqref{qpscDef}, and the
functional $\fF_{\omega_T,t}^\anc$ need not be the Laplace transform of a
probability measure. As we shall see in Section~\ref{ssec:prefeast} it will also
be less invasive than the two-time measurement.

\subsection{Comparing the entropic functionals}
\label{ssec: quantumfunct comparison}


In the previous sections we have introduced several approaches to study entropic
fluctuations (naive, 2TM, QPSC and EAST), and we shall give a first comparison
between them.

For infinitely extended systems the naive approach of Section~\ref{ssec:naive},
with $c^t$ defined by~\eqref{eq:ctDef}, would lead to the following functionals 
(compare with~\eqref{eq:naivelaplace})
\[
\fF_{\omega_T,t}^\mathrm{naive}(\alpha)\coloneq \omega_T\left(\e^{-\alpha c^t}\right),
\]
while the 2TM, QPSC and EAST functionals are respectively given
by~\eqref{def:ttmextended}, \eqref{qpscDef} and~\eqref{def:ancillainfinite}.

When $T$ is finite $\omega_T$ is $\omega$-normal and the different functionals
can be further rewritten in the GNS representation. Namely,
\begin{align*}
\fF_{\omega_T,t}^\mathrm{naive}(\alpha)&=\left\bra\Omega_T, \e^{-\alpha c^t} \Omega_T  \right\ket,\\[1mm]
\fF_{\omega_T,t}^\ttm(\alpha)&=\lim_{R\to\infty}\frac1R\int_0^R  \left\bra \Omega_T , \Delta_\omega^{i\theta} \e^{-\alpha c^t+\alpha \log\Delta_\omega} \e^{-\alpha \log \Delta_\omega} \Delta_\omega^{-i\theta} \Omega_T \right\ket \,  \d\theta, \\[6pt]
\fF_{\omega_T,t}^\qpsc(\alpha)&=\left\bra\Omega_T, \e^{-\alpha c^t+\alpha \log\Delta_\omega} \e^{-\alpha \log \Delta_\omega}\Omega_T \right\ket, \\[6pt]
\fF_{\omega_T,t}^\anc(\alpha)&=\left\bra\Omega_T, \e^{-\frac{\alpha}{2} \log \Delta_\omega} \e^{-\alpha c^t+\alpha \log\Delta_\omega} \e^{-\frac{\alpha}{2} \log \Delta_\omega}\Omega_T \right\ket.
\end{align*}
In the classical setting all these quantities coincide due to the triviality of
$\Delta_\omega$, and are given by~\eqref{eq:classicallaplace}. In other words,
the various approaches are indistinguishable in the classical/commutative
framework, and the naive, 2TM, QPSC and EAST functionals discussed above are
natural non-commutative extensions of the classical theory. The diversity of
non-commutative/quantum theory of entropic fluctuations then stems from the non
triviality and the rich KMS structure associated to $\omega$ via
$\Delta_\omega$.  In fact one can construct an entire host of mathematically
natural non-commutative extensions of $\fF_{\omega_T,t}$;
see~\cite[Section~3.3]{Jaksic2010b}. Some of these non-commutative extensions
have direct quantum mechanical interpretations, like the 2TM and EAST ones, and
some do not, or such interpretations are not known at the moment. For example,
if $\nu\neq\omega$, then to the best of our knowledge, the functional
$\fF_{\nu,t}^\qpsc$ does not have any quantum mechanical or stochastic
interpretation.

\section{The quantum principle of regular entropic fluctuations (PREF)}
\label{sec:quantumpref}

In this section we elaborate on the classical PREF described at the end of 
Section~\ref{sec:classic}.

\subsection{Definition of the quantum PREF}

As we have already stressed, while the 2TM functionals are Fourier transforms of
probability measures, the EAST and QPSC functionals need not have this property
(at $T=0$ the three functionals coincide). Hence, an associated LDP is not
available in general. Concerning the EAST entropic functionals, when $T=0$ they
provide natural operational access to the 2TM functional. The question of
their connection for later time, eventually in the NESS, is therefore natural.
Moreover, as mentioned in the introduction and in view of the tomography part of
the EAST protocol, in this indirect measurement approach one needs to repeat the
experiment with the \textsl{same} ``initial'' state, \ie the state of the open
quantum system at the time when it is coupled to the ancilla. In this respect
the NESS is a natural choice putting emphasis on $\fF_{\omega_\infty,t}^\anc$.
Concerning the QPSC approach, its close parallel with the classical setting
makes it natural to incorporate it in the quantum version of PREF which we
define below, compare with~\eqref{def:GEfunction} and~\eqref{eq:classicalpref}
in Section~\ref{sec:classic}.

The quantum version of PREF given in Definition~\ref{def-PREF} below, requires
extending the various entropic functionals to $\alpha$'s in a real interval
(they are up to now defined only for $\alpha\in\i\RR$). This is possible under
the following assumption ($\vartheta$ will be specified later on).
\assuming{AnV$\boldsymbol{(\vartheta)}$}{AnV}{%
The connecting perturbation $V$ is such that the map
$
\i\RR\ni\theta\mapsto\varsigma_\omega^{-\i\theta}(V)\in\cO
$
has an analytic extension to the vertical strip 
$\fS(\vartheta)=\left\{ z\in \CC\mid|\Re\, z| <\vartheta\right\}$.
}
If Assumption~\AnV{} holds then, for any $t\in\RR$, the function
\[
\i \RR\ni \alpha \mapsto [D\omega_{t}: D\omega]_\alpha \in \cO
\]
has an analytic extension to $\fS(\vartheta)$, and hence so do the various
entropic functionals.\footnote{The EAST functional actually extends to the strip
$\fS(2\vartheta)$.} At this point, the sole analytic extension of $[D\omega_{t}:
D\omega]_\alpha$ would be sufficient, see~\cite{Benoist2025a}. Since we will
need the stronger Assumption~\AnV{} later on,  for the sake  of  simplicity in
the exposition we assume it holds throughout this section.

We also assume \NESS{} and ergodicity of the reservoir systems, so that the NESS
2TM functional is defined as above. We use the formulation of PREF
in~\cite{Benoist2026}, which adds the QPSC equality to the strong PREF
of~\cite{Benoist2025a}. In the latter work, weak PREF means that the transient
and steady-state 2TM statistics satisfy large deviation principles with the same
rate function.

\begin{definition}\label{def-PREF} Let $I=\,]\vartheta_-,\vartheta_+[$ be an
open interval containing $0$ and contained in $]-\vartheta,\vartheta[$. We say
that $(\cO,\tau,\omega)$ satisfies the PREF on the interval $I$ if, for all
$\alpha\in I$, the limits
\beq
\begin{split}
\bF^\ttm_{\omega}(\alpha)&=\lim_{t\to\infty}\frac1t\log\fF_{\omega,t}^\ttm(\alpha),\\[1mm]
\bF^\ttm_{\omega_\infty}(\alpha)&=\lim_{t\to\infty}\frac1t\log\fF_{\omega_\infty,t}^\ttm(\alpha),\\[1mm]
\bF^\anc_{\omega_\infty}(\alpha)&=\lim_{t\to\infty}\frac1t\log\fF^\anc_{\omega_\infty,t}(\alpha),\\[1mm]
\bF^\qpsc_{\omega_\infty}(\alpha)&=\lim_{t\to\infty}\frac1t\log\fF^\qpsc_{\omega_\infty,t}(\alpha)
\end{split}
\label{def:functionals-limit}
\eeq
exist, and define differentiable functions on $I$, satisfying
\beq
\bF^\ttm_{\omega}=\bF^\ttm_{\omega_\infty}=\bF^\anc_{\omega_\infty}=\bF^\qpsc_{\omega_\infty}.
\label{def:pref}
\eeq
For the QPSC functional, the logarithm is understood as the principal value.  
We denote by $\bF$ the common function in~\eqref{def:pref}.
\end{definition}

\subsection{PREF and 2TM}

Using the Gärtner--Ellis Theorem, if the limit
\[
\bF^\ttm_{\omega}(\alpha)\coloneq\lim_{t\to\infty}\tfrac1t\log\fF_{\omega, t}^\ttm(\alpha)
\]
exists for all $\alpha\in]\vartheta_-,\vartheta_+[$, with
$\max\{|\vartheta_-|,\vartheta_+\}<\vartheta$, and defines a differentiable
function on this interval, then the family $(P_{\omega,t}^\ttm)_t$ satisfies the
restricted LDP on the interval of slopes specified by the Gärtner--Ellis Theorem
in Section~\ref{sec:classic}.  Under the assumptions of
Theorem~\ref{thm:stability}, when $\omega_{T,\sS}$ is faithful it is an
immediate consequence of Equation~\eqref{eq:ttmstability} that all the limits
\[
\bF^\ttm_{\omega_T}(\alpha)\coloneq \lim_{t\to\infty}\tfrac1t\log\fF_{\omega_T, t}^\ttm(\alpha)
\]
then exist and coincide. This is always the case for finite $T$, and when 
$\omega_{\infty,\sS}$ is faithful we get
\[
\bF^\ttm_{\omega} = \bF^\ttm_{\omega_\infty}.
\]

In the 2TM approach, the existence and differentiability of the first, respectively
second limit in~\eqref{def:functionals-limit} imply the corresponding
restricted LDP by the Gärtner--Ellis Theorem; a full LDP requires its additional
global hypotheses. Indeed, as in the classical case, consider the families of
probability measures $(P_{\omega_T,t}^\ttm)_{t>0}$, $T\in\R_+\cup\{\infty\}$,
defined by
\[
P_{\omega_T,t}^\ttm(B)=Q_{\omega_T,t}^\ttm(tB),
\]
for all Borel sets $B\subset\RR$ and $t>0$. They describe the statistics of the
two-time measurement entropy production per unit time for a protocol initialized
in the state \(\omega_T\), with time \(t\) between the two measurements. The
following results are then immediate consequences of
Proposition~\ref{prop:ttmtransientinfinite} and Theorem~\ref{thm:stability}
respectively.
\begin{enumerate}
\item If $(P_{\omega,t}^\ttm)_t$ satisfies a full LDP, \ie  a quantum Evans--Searles 
fluctuation theorem holds, with rate function $\II$ and the system is TRI, 
then for all $s\in \RR$,
\beq
\II(-s)= \II(s) + s,
\label{quantum-es-1}
\eeq
\ie a quantum Evans--Searles fluctuation relation also holds. If the restricted 
LDP holds on $]-a,a[$ for some $a>0$, then~\eqref{quantum-es-1} holds for $s\in]-a,a[$.
\item Under the assumptions of Theorem~\ref{thm:stability}, if the family
$(P_{\omega,t}^\ttm)_t$ satisfies a (full or restricted) LDP then each family
$(P_{\omega_T,t}^\ttm)_t$ with faithful $\omega_{T,\sS}$ satisfies the same LDP,
and with the \textsl{same rate function}. In particular, this applies at
$T=\infty$ when $\omega_{\infty,\sS}$ is faithful. Under these hypotheses, in
the 2TM approach the validity of the Evans--Searles fluctuation theorem, with
the associated Evans--Searles fluctuation relation when the system is TRI,
automatically guarantees the validity of a quantum Gallavotti--Cohen fluctuation
theorem and relation. This is in stark contrast with the classical situation.
\end{enumerate}

\subsection{PREF and EAST}
\label{ssec:prefeast}

While in the classical case PREF can be regarded as a deep dynamical property of
the system exemplified by the exchange of limits~\eqref{cl-ex-lim-1}, it
essentially trivializes in the 2TM approach because of the first measurement's
dominating effect in the thermodynamic limit.  In a surprising turn, the
parallel with the classical PREF is restored when considering the EAST
functionals.  Indeed, if Assumption~\AnV{} holds with $\vartheta>\frac12$, for
any real $\alpha\in]-\vartheta,\vartheta[$ and finite $T$ we actually have the
following estimate, see~\cite[Proposition 2.12]{Benoist2025a},
$$
\e^{-2|T|(\|\varsigma_\omega^{\i/2}(V)\|+\|V\|)}\fF_{\omega, t}^\ttm(\alpha)
\leq\fF_{\omega_T, t}^\anc(\alpha)\leq \e^{2|T|(\|\varsigma_\omega^{-\i/2}(V)\|+\|V\|)}
\fF_{\omega, t}^\ttm(\alpha).
$$
For any finite time $T$, assuming the existence of the first limit
in~\eqref{def:functionals-limit} this gives that, for
$\alpha\in]\vartheta_-,\vartheta_+[$ and \textsl{arbitrary} finite initial time
$T$,
\beq\label{eq:finitetimeeastpref}
\bF_{\omega_T}^\anc(\alpha)=\bF_{\omega}^\ttm(\alpha).
\eeq
In other words, as mentioned in the introduction, in great generality the EAST
protocol satisfies a weaker form of PREF, valid for each fixed finite waiting
time $T$.

However, if one wants to go up to $T=\infty$, \ie  consider the NESS
$\omega_\infty$, under the hypotheses ensuring
$\bF_\omega^\ttm=\bF_{\omega_\infty}^\ttm$, the second equality
in~\eqref{def:pref} holds if and only if the exchange of limits, compare
with~\eqref{cl-ex-lim-1},
$$
\lim_{T\to\infty}\lim_{t\to\infty}\tfrac1t\log\fF_{\omega_T,t}^\anc(\alpha) 
=\lim_{t\to\infty}\lim_{T\to\infty}\tfrac1t\log\fF_{\omega_T, t}^\anc(\alpha)
$$
is valid for $\alpha\in]\vartheta_-, \vartheta_+[$.

\section{Quantum transfer operators approach to PREF}
\label{sec:quantumtransferoperator}

In this section, we briefly present the quantum transfer operators
representation of the various entropic functionals, and we provide a
characterization of the quantum PREF in terms of the spectral resonance of these
operators. This approach builds on the quantum transfer operator construction of
NESS developed in~\cite{Jaksic2002a}; see also~\cite{Jaksic2010b} for a
pedagogical introduction to the topic. We then show how it can be applied to one
of the paradigmatic models of quantum statistical mechanics, namely the
spin-fermion model.

\subsection{Liouvillean representation of entropic functionals}

Again we suppose that Assumption~\AnV{} holds for some $\vartheta>1/2$.
For all $\alpha\in\fS(\vartheta)$, we define the following families of
so-called $\alpha$-Liouvilleans:
\beq\label{def:alphaliouv}
\begin{split}
\cL_{\alpha} & = \cL_\free+V-J\varsigma_\omega^{-\i\bar \alpha}(V)J,\\[1mm]
\widehat\cL_\alpha & = \cL_\free+\varsigma_\omega^{\i(1/4-\alpha/2)}(V)
-J\varsigma_\omega^{-\i(1/4+\bar \alpha/2)}(V)J.
\end{split}
\eeq
Here $\cL_\free$ is the standard Liouvillean of the non-interacting open
system, $J$ is the modular conjugation and $\varsigma_\omega$ the modular
group (see Section~\ref{ssec:modular}).

\noindent\textbf{Remark.} The $\cL_\alpha$'s differ from the standard
Liouvillean $\cL$ only by a term in the commutant $\fM'$, namely
\[
\cL_{\alpha} = \cL+J(V-\varsigma_\omega^{-\i\bar \alpha}(V))J,
\]
and hence also implement the dynamics, \ie
$\e^{\i t\cL_\alpha}A\e^{-\i t\cL_\alpha}=\tau^t(A)$ for all $A\in\fM$.

\medskip
The primary reason for introducing these $\alpha$-Liouvilleans is the
following representation of the entropic functionals~\cite[Proposition~3.2
and Equations~(5.11)--(5.12)]{Benoist2025a}.
\bep\label{prop:liouv-functionals} For all $\alpha\in\i\RR$ and $t,T\in\RR$,
\ben
\item $\ds\fF_{\omega,t}^\ttm(\alpha)=\langle\Omega,\e^{\i t\cL_{1/2-\alpha}}\Omega\rangle$.
\item $\ds\fF_{\omega_T,t}^\qpsc(\alpha)=\langle\Omega,\e^{\i T\cL_{1/2}}[D\omega_{-t}:D\omega]_\alpha\Omega\rangle
=\langle\Omega,\e^{\i T\cL_{1/2}}\e^{\i t\cL_{1/2-\alpha}}\Omega\rangle$.
\item $\ds\fF_{\omega_T,t}^\anc(\alpha)=\langle\Omega,\e^{\i T\cL_{1/2}}[D\omega_{-t}:D\omega]^\ast_{\frac{\bar\alpha}{2}}[D\omega_{-t}:D\omega]_{\frac{\alpha}{2}}\Omega\rangle
=\langle\Omega,\e^{\i T\cL_{1/2}}\e^{\i t\widehat\cL_{1/2-\alpha}}\Omega\rangle$.
\een
The identity in~(1) analytically extends to the strip
$|\Re(\alpha-\frac12)|<\vartheta$, the one in~(2) extends to the strip
$\frac12-\vartheta<\Re(\alpha)<\vartheta$, while the one in~(3) extends
to the strip $1-2\vartheta<\Re(\alpha)<2\vartheta$. We note that the first
and last strips contain the real interval $[0,1]$.
\eep

Based on this representation, the study of the limits
in~\eqref{def:functionals-limit} is linked to the large $T$, then $t$,
behavior of the semigroups generated by the $\alpha$-Liouvilleans.
This behavior can be studied via spectral resonance theory, as we now
briefly explain. Because $\cL_\alpha$ is a bounded perturbation of the
self-adjoint operator $\cL_\free$, its resolvent
$z\mapsto(z-\cL_\alpha)^{-1}$ is well-defined on some lower half-plane
$\Im(z)<-m_\alpha$. Suppose that, between suitable vectors, it has a
meromorphic continuation to $\Im(z)<\mu_\alpha$, with $\mu_\alpha>0$,
whose only singularity is a simple pole $\cE(\alpha)$. Under the
resolvent bounds stated in~\cite[Proposition~4.1]{Benoist2025a}, one
obtains, for $\max\{0,\Im\cE(\alpha)\}<\gamma<\mu_\alpha$,
\beq\label{eq:resonancelimit1}
\bra\phi,\e^{\i t\cL_\alpha}\psi\ket
=\e^{\i t\cE(\alpha)}\bra\phi,P_\alpha\psi\ket
+O\left(\e^{-\gamma t}\right),
\eeq
where $\bra\phi,P_\alpha\psi\ket$ denotes the residue of the continued
scalar resolvent. If this coefficient is nonzero and $\cE(\alpha)$ is
purely imaginary, as for the real parameters relevant below, then
\beq\label{eq:resonancelimit2}
\lim_{t\to\infty}\frac1t\log\bra\phi,\e^{\i t\cL_\alpha}\psi\ket
=\i\cE(\alpha).
\eeq

The analysis of the large $T$ limit is motivated by the spectral theory
of NESS developed in~\cite{Jaksic2002a}. When the spectral approach
applies to both limits, it gives that $\bF_\omega^\ttm(\alpha)$ and
$\bF_{\omega_\infty}^\qpsc(\alpha)$ exist and equal
$\i\cE\left(\frac12-\alpha\right)$, while
$\bF_{\omega_\infty}^\anc(\alpha)=\i\hat\cE\left(\frac12-\alpha\right)$,
where $\hat\cE(\alpha)$ is the corresponding pole for
$\widehat\cL_\alpha$. The two families are closely related: on
appropriate domains, their definitions give
\[
\widehat\cL_\alpha
=\Delta_\omega^{\alpha/2-1/4}\cL_\alpha\Delta_\omega^{-\alpha/2+1/4}.
\]
The exact equality $\fF_{\omega,t}^\anc=\fF_{\omega,t}^\ttm$ identifies
their leading resonances once the corresponding expansions have been
established with nonzero coefficients. Together with differentiability
in $\alpha$ and the 2TM stability result, this yields a model-independent
spectral criterion for the PREF.

The difficulty of this spectral approach comes from the fact that the
meromorphic continuation is typically available only on a suitable
subspace: one studies the map $z\mapsto D(z-\cL_\alpha)^{-1}D$ for some
bounded operator $D\geq0$ with dense range. The
relations~\eqref{eq:resonancelimit1}--\eqref{eq:resonancelimit2} then
apply to vectors $\phi,\psi\in\ran(D)$. Such continuations are obtained
via spectral deformation techniques initially developed in the theory
of Schrödinger operators. This parallels the Ruelle transfer operator
approach to classical dynamical systems, for which one has to consider
suitable spaces of functions, e.g. Hölder continuous function spaces,
to unveil the so-called Ruelle--Pollicott resonances.

Finally, we mention that the combination of the large $T$ and $t$
limits makes this approach quite technical.
In~\cite[Section~4.3]{Benoist2025a} a general axiomatic scheme is
proposed that allows one to derive the PREF using this spectral
resonance method. As expected given the information it yields, this
scheme is technically involved to apply to concrete physically relevant
models. In~\cite{Benoist2026} we apply it to generalized versions of the
spin--fermion model, where a finite level system is coupled to several
fermionic free reservoirs with interactions which are polynomial in the
annihilation/creation operators. In the next section we show how this
can be used in the context of the usual spin--fermion model.

\subsection{Spin Fermion (simple version)}
\label{ssec:spinfermion}

The spin--fermion model is a concrete example of an open quantum system with the
structure described in Section~\ref{ssec:oqs}. In its simplest version $\sS$ is
a two-level system, \ie $\cK_\sS=\CC^2$, with Hamiltonian $H_\sS=\sigma_z$. Each
reservoir subsystem $\sR_j$ is a free Fermi gas with single particle Hilbert
space $\fh_j$, single particle Hamiltonian $h_j$ and initial KMS state at
inverse temperature $\beta_j$, see Example 2 in Section~\ref{ssec:cstar}.

For each $j$, the interaction of $\sS$ with $\sR_j$ is of the form
\[
V_j = \sigma_x \otimes \varphi_j(f_j),
\]
where $\varphi_j(f)=\frac1{\sqrt2}(a_j(f)+a_j^\ast(f))$ is the Segal field
operator and $f_j\in \fh_j$ is a form factor. The complete interaction is
$V=\lambda(V_1+V_2)$ where $\lambda\in\RR$ is a coupling constant.

The following technical assumptions allow one to apply the Liouvillean approach
presented in the previous section.

\assuming{SFM1}{SFMone}{
The single particle spaces $\fh_j$ are of the form $\fh_j=L^2(\RR_+,\d
s)\otimes\fH$ for some Hilbert space $\fH$ equipped with a complex conjugation
$\overline{\,\cdot\vphantom{X}\;}$, and $h_j$ is the operator of multiplication
by the variable $s\in\RR_+$. } For $j=1,2$ we associate to the form factor $f_j$
the function $\widetilde{f_j}\in L^2(\RR,\d s)\otimes\fH$ defined as
$$
\widetilde{f_j}(s)=
\begin{cases}
f_j(s)&\text{if }s\geq0;\\[1mm]
\BAR{f_j(|s|)}&\text{if }s<0.
\end{cases}
$$
and assume
\assuming{SFM2}{SFMtwo}{
For every $r>0$, $\widetilde{f_j}$ is in the Hardy class $H^2(r)$ of all analytic functions 
\[f:\{z\in\CC\mid |\Im z|<r\}\to\fH\] such that
\[
\|f\|^2_{H^2(r)}=\sup_{|\theta|<r}\int_\RR\|f(s+\i\theta)\|_{\fH}^2\d s<\infty,
\]
and, for all $r>0$ and all $a>0$ one has
\[
\sup_{|\theta|<r}\int_\RR\e^{a|s|}\|\widetilde{f_j}(s+\i\theta)\|_\fH^2\d s<\infty.
\]
}
\assuming{SFM3}{SFMthree}{
$\|\tilde f_j(2)\|^2_{\fH}>0$.
}
\assuming{SFM4}{SFMfour}{
$f_j=\BAR{f_j}$.
}
 
Assumption~\SFMone{} ensures that the reservoirs are ergodic, while
\SFMtwo{} provides the regularity needed for the analytic deformation
of the Liouvilleans. Assumption~\SFMthree{} is the Fermi golden rule
condition: it requires nonzero coupling to each reservoir at the spin
transition energy $2$. Together these assumptions allow one to apply
the spectral approach for sufficiently small nonzero coupling.
Finally, \SFMfour{} ensures that the system is TRI. Note that the generic
assumptions~\Reg{} and~\AnV{} translate respectively to
$f_j\in Dom(h_j)$ and $f_j\in Dom(e^{\pm\beta_j\theta h_j})$,
and thus follow from~\SFMtwo{}. 

The following result was proven in~\cite{Benoist2026}.
\begin{theorem}\label{main-thm-sfm} Suppose that~\SFMone--\SFMfour{} hold. Then:
\ben
\item There exists $\Lambda >0$ such that for $0<|\lambda|<\Lambda$, \NESS holds 
and the restriction $\omega_{\infty,\sS}$ of the NESS $\omega_\infty$ to
$\cO_\sS$ is faithful.
\item For any $\vartheta >0$ there exists $\Lambda >0$ such that  the PREF holds 
on $]-\vartheta, 1+\vartheta[$ for $0 <|\lambda|<\Lambda$. Moreover, the function
\[
]-\vartheta, 1+\vartheta[\, \ni \alpha \mapsto  \bF(\alpha)
\]
is real-analytic and satisfies the symmetry $\bF(\alpha)=\bF(1-\alpha)$. It is 
identically vanishing on $]-\vartheta, 1+\vartheta[$ if $\beta_1=\beta_2$, and 
otherwise strictly convex on this interval.
\item For small $\lambda$, one has
\[
\bF(\alpha)=\lambda^2\bF_2(\alpha)+O(\lambda^3),
\]
where
\begin{align*}
\bF_2(\alpha)=-\frac{\pi}2\left(\vphantom{\sqrt{\sum_{j,k=1}^2}} \sum_{j=1}^2\right.&\|\tilde f_j(2)\|^2_{\fH}\\
&\left.-\sqrt{\sum_{j,k=1}^2 \left(\tanh(\beta_j)\tanh (\beta_k) +\frac{\cosh((\beta_j-\beta_k)(1-2\alpha))}{\cosh(\beta_j)\cosh(\beta_k)}\right) \|\tilde f_j(2)\|^2_{\fH}\|\tilde f_k(2)\|^2_{\fH}}\,\right).
\end{align*}
\een
\end{theorem} 

\noindent\textbf{Remark.} The fact that $\omega_{\infty,\sS}$ is faithful
guarantees that the stability Theorem~\ref{thm:stability} can be applied to the
NESS $\omega_\infty$. Moreover, $-\bF^\prime(0)=\omega_\infty(\sigma)$ is the
mean steady-state entropy production rate. It is positive when
$\beta_1\neq\beta_2$ and vanishes at equilibrium.

\bigskip
{\small\textbf{Data availability.} No data is associated with this work.}

{\small\textbf{Conflict of Interest.} The Authors have no COI pertaining to this work.}

\bibliographystyle{capalpha}
\bibliography{proceedings}

\end{document}